\documentclass[a4paper,11pt]{article}
\usepackage{jheppub} 
\usepackage{lineno}
\usepackage{amsfonts}
\usepackage{mathrsfs}
\usepackage{bbold}
\usepackage{subcaption}
\usepackage[dvipsnames]{xcolor}
\usepackage{mathtools}
\usepackage[T1]{fontenc}
\usepackage{amssymb,ascmac,fancybox,mathrsfs}
\usepackage{color,booktabs,fancyhdr,bm,comment,cancel}
\usepackage{hyperref,appendix}
\usepackage{graphicx,graphics}

\newcommand{\thalf}{{\tfrac{1}{2}}}

\DeclareMathOperator{\Tr}{\mathrm{Tr}}

\title{Quasinormal modes of a Proca field in Schwarzschild-AdS$_5$ spacetime via the isomonodromy method}

\author[a]{Juli\'{a}n Barragán Amado,}
\author[b]{Tiago V. Fernandes}
\author[b]{and David C. Lopes}

\affiliation[a]{Grupo de F\'{i}sica Matem\'{a}tica,
Departamento de Matem\'{a}ica, Instituto Superior T\'{e}cnico, Universidade de Lisboa, Avenida Rovisco Pais 1, 1049-001 Lisboa, Portugal}

\affiliation[b]{Centro de Astrof\'isica e Gravita\c c\~ao 
- CENTRA, Departamento de F\'isica, Instituto Superior T\'{e}cnico 
- IST, Universidade de Lisboa - UL,
 Avenida Rovisco Pais 1, 1049-001 Lisboa, Portugal}

\emailAdd{jose.barragan.amado@tecnico.ulisboa.pt}
\emailAdd{tiago.vasques.fernandes@tecnico.ulisboa.pt}
\emailAdd{david.d.lopes@tecnico.ulisboa.pt}

\abstract{We consider Proca field perturbations in a five-dimensional Schwarzschild-anti--de Sitter (Schwarzschild-AdS$_{5}$) black hole geometry. Using the vector spherical harmonic (VSH) method, we show that the Proca field decomposes into scalar-type and vector-type components according to their tensorial behavior on the three-sphere. Two degrees of freedom of the field are described by scalar-type components, which are coupled due to the mass term, while the remaining two degrees of freedom are described by a vector-type component, which decouples completely. Motivated by the Frolov-Krtou\v{s}-Kubiz\v{n}\'{a}k-Santos (FKKS) ansatz in the limit of zero spin, we use a field transformation to decouple the scalar-type components at the expense of introducing a complex separation parameter $\beta$. This parameter can be determined analytically, and its values correspond to two distinct polarizations of the scalar-type sector: ``electromagnetic'' and ``non-electromagnetic'', denoted by $\beta_{+}$ and $\beta_{-}$, respectively. In the scalar-type sector, the radial differential equation for each polarization is a Fuchsian differential equation with five singularities, whereas in the vector-type sector, the radial equation has four singularities. By means of the isomonodromy method, we reformulate the boundary value problem in terms of the initial conditions of the Painlev\'{e} VI $\tau$ function and, 
using a series expansion of the $\tau$ function, we compute the scalar-type and vector-type quasinormal modes (QNMs) in the small horizon limit.
Our results are in overall very good agreement with those obtained via the numerical integration method. This shows that the 
isomonodromy method is a reliable method to compute quasinormal modes in the 
small horizon limit with high accuracy.}

\makeatletter\gdef\@fpheader{}\makeatother

\begin{document}
\maketitle
\flushbottom

\section{Introduction}
\label{sec:1}

Black hole spacetimes are encompassed in the plethora of stationary solutions 
of the Einstein equations. The interest in these spacetimes is vast as they 
can model very compact objects residing at the center of galaxies, for example.
It is thus important to understand their stability. By performing linear
perturbations of existing fields and of spacetime itself, black holes 
respond by vibrating through the quasinormal modes (QNMs) at intermediate times.
Quasinormal modes are 
perturbations characterized by a complex frequency, where the real 
part describes the oscillation of the perturbations, and the 
imaginary part describes the decay, if negative, or growth, if positive, of the 
perturbations. For the spacetime to be stable to linear perturbations, 
the imaginary part must thus be negative for all the quasinormal modes. 

The study of quasinormal modes in AdS spacetimes has been of interest, 
namely due to the AdS/CFT correspondence \cite{Maldacena:1997re}. This 
correspondence establishes that asymptotically AdS black branes, which 
can be described approximately by very large black holes, are dual to 
a thermal state of a conformal field theory in less one dimension. Moreover, 
the quasinormal modes of these spacetimes are dual to the response 
of the thermal state under perturbations in the conformal field theory side \cite{Witten:1998zw, Horowitz:1999jd}.

There has been an extensive study of quasinormal modes of different fields with 
different spins. The quasinormal modes of a scalar field in $d$-dimensional 
Schwarzschild-AdS has been studied in \cite{Horowitz:1999jd}, while 
electromagnetic and gravitational quasinormal modes and their asymptotic 
behaviour were investigated 
in four dimensional Schwarzschild-AdS in \cite{Cardoso:2001bb,Cardoso:2003cj}.
The latter study used Dirichlet boundary conditions for the electromagnetic 
field, while other boundary conditions were explored in \cite{Wang:2015goa}. 
For higher dimensions in spherically symmetric spacetimes, the 
equations that govern the perturbations can be obtained using the 
Kodama and Ishibashi's decomposition~\cite{Kodama:2003jz,Ishibashi:2003ap,
Kodama:2003kk}, or as we call it here, the vector spherical harmonics (VSH) method. 
The master equation and its eigenvalues in pure AdS were 
studied in \cite{Ishibashi:2004wx}.
The equations for the electromagnetic field were derived for higher 
dimensions in \cite{Crispino:2000jx} to compute absorption cross sections 
of electromagnetic radiation, and in \cite{Lopez-Ortega:2006vjp} to obtain
the asymptotic quasinormal modes of the electromagnetic field in a 
$d$-dimensional Schwarzschild-AdS black hole.

For the case of a massive vector field, i.e. a Proca field, the normal modes were computed analytically for pure AdS in four dimensions in \cite{Fernandes:2022con}. 
The quasinormal modes for Schwarzschild and Schwarzschild-AdS in four-dimensions were analyzed in \cite{Rosa:2011my} and \cite{Konoplya:2005hr, Fernandes:2021qvr}, respectively.
In higher dimensions, the normal modes in pure AdS were computed in 
\cite{Lopes:2024ofy}. The Proca equations in a $d$ dimensional Schwarzschild spacetime were derived in \cite{Herdeiro:2011uu} and the perturbation expansion was obtained in \cite{Ueda:2018xvl} for near extremal black hole spacetimes. The quasinormal modes for the Proca field in higher dimensions were also studied in~\cite{Lopes:2025}.

The 
Proca field in a $d$-dimensional spacetime 
possesses $d-1$ degrees of freedom, with $d-3$ degrees 
of freedom in the vector-type sector where the equations are decoupled, 
and two degrees of freedom in the scalar-type sector, where the 
equations are coupled. The fact that the scalar-type sector is coupled 
does not prevent the use of numerical methods to obtain the quasinormal modes. 
Namely in the Schwarzschild-AdS spacetime, 
one can use either the Horowitz-Hubeny method 
\cite{Horowitz:1999jd,Fernandes:2021qvr,Delsate:2011qp} or numerical integration of 
the equations \cite{Chandra:1975, Konoplya:2008rq, Zhidenko:2009zx, Pani:2013pma}. 
These numerical methods converge effectively for large black holes, with the 
numerical integration method having better results and convergence for intermediate 
black holes. However, for small black hole radius, the methods lose accuracy. 
Having a coupled system also contributes to the 
loss of accuracy of the numerical method. One way of seeing this is that the method involves finding the root of a matrix determinant, which becomes less accurate for increasing dimensions of the matrix in the occurrence of a badly conditioned matrix. 
Hence, the decoupling of the scalar-type sector is thus of 
great importance both to compute the QNMs more accurately and to obtain 
a physical interpretation of the two degrees of freedom at play.

The separation and decoupling of the Proca field in general Kerr-NUT-(A)dS 
spacetimes~\cite{Chen:2006xh} has been 
accomplished with the Frolov-Krtou\v{s}-Kubiz\v{n}ák-Santos (FKKS) ansatz
~\cite{Frolov:2018ezx}, motivated by the work of Lunin~\cite{Lunin:2017drx}. 
The ansatz was developed using the hidden symmetries of these stationary 
spacetimes generated by a closed conformal Killing-Yano 
$2$-form~\cite{Frolov:2017kze}. In the Schwarzschild limit, this ansatz 
gives a transformation that enables the decoupling of the scalar-type sector, 
obtained in \cite{Percival:2020skc} for four dimensional Schwarzschild and 
in~\cite{Fernandes:2021qvr} for four dimensional Schwarzschild-AdS. 
To our knowledge, the corresponding transformation for five dimensions has not been achieved.


After the separation of variables and the decoupling of the system, the equations for the field perturbations typically reduce to second-order ordinary differential equations (ODEs) characterized by a fixed number of singular points and can be written as Heun-type equations. In recent years, the isomonodromy method has been introduced to study these differential equations. In particular, the correspondence between a Fuchsian system with four singular points and its monodromy representation can be described through a set of equations satisfying a zero-curvature condition while preserving its monodromy data. Notably, these isomonodromic deformation equations reduce to the Painlev\'{e} VI (PVI) equation \cite{Schlesinger:1912}. Garnier demonstrated that the isomonodromic equations can be framed as a Hamiltonian system that evolves according to the PVI Hamiltonian \cite{Garnier:1912}.  Building on this, Jimbo, Miwa, and Ueno introduced the isomonodromic $\tau$ function as the generating function of the isomonodromic Hamiltonian, which generates the isomonodromic flow \cite{Jimbo:1981tov,Jimbo:1981zz}. By specifying initial conditions for this flow, we are led to a set of transcendental equations involving the PVI $\tau$ function discovered by Gamayun et al. \cite{Gamayun:2012ma,Gamayun:2013auu}, based on earlier work by Jimbo \cite{Jimbo:1982}. The latter establishes a mapping that determines the accessory parameter of the Heun equation in terms of the monodromies of the Fuchsian system \cite{Novaes:2014lha,CarneirodaCunha:2015qln}.  
In this regard, the isomonodromic $\tau$ functions of Painlevé transcendents have been applied to describe various physical systems, including the Rabi model in quantum optics \cite{CarneirodaCunha:2015vxu}, conformal maps of polycircular arc domains \cite{Anselmo:2018zre,Anselmo:2020bmt,CarneirodaCunha:2021jsu}, and the computation of QNMs in different backgrounds \cite{Novaes:2018fry,BarraganAmado:2018zpa,daCunha:2021jkm,Cavalcante:2021scq,Amado:2021erf}. More recently, the connection between the monodromy parameters and the confluent Heun equation in the case of massive scalar perturbations in Kerr black holes has shed light on the existence of a geometric phase around a point of degeneracy, where the fundamental QNM and its first overtone coincide \cite{Cavalcante:2024swt,Cavalcante:2024kmy}.

In this work, we use the isomonodromy method to compute the quasinormal modes of a Proca field in a Schwarzschild-AdS$_{5}$ spacetime, for small to intermediate black holes.  As this method requires decoupled equations, we find a transformation motivated by the FKKS ansatz to separate the scalar-type sector of the Proca field. In the scalar-type sector, we find that the radial differential equation for each polarization is a second-order Fuchsian differential equation with five singularities, whereas in the vector-type sector, the radial equation has four singularities. With regard to the number and character of the singular points, the resulting radial ODEs are Heun-type equations. The isomonodromic $\tau$ function can then be used to obtain the quasinormal modes of the Proca field for each sector and polarization. 
It is expected that this method gives the eigenfrequencies more accurately than the numerical integration method for small black holes. A comparison is made between these methods. It is found that both methods are in overall good agreement, where the largest differences occur for the imaginary part at small horizon radii and at $r_h$ close to unity. Although the exact value of the quasinormal modes is not known, 
the comparison agrees with the expectation that the isomonodromy method is a viable method to compute the quasinormal modes with high accuracy for small horizon radius.

The paper is organized as follows. In Sec.~\ref{sec:2}, we present the  Einstein-Proca field equations for a fixed background geometry with a negative cosmological constant. In Sec.~\ref{sec:3}, we study the separation and decoupling of the Proca field in the Schwarzschild-AdS$_{5}$ black hole. First, by applying the VSH method, we separate the scalar-type and vector-type components of the Proca field. Then, using the FKKS method, we decouple the scalar-type components at the expense of introducing a separation parameter $\beta$. We present the radial differential equations for each sector, which can be written as Heun-like equations. In Sec.~\ref{sec:4}, we apply the isomonodromy method to recast the boundary value problem of the scalar-type and vector-type in terms of the initial conditions of the Painlev\'{e} VI $\tau$ function. In Sec.~\ref{sec:5}, we compute the quasinormal modes as a function of the event horizon radius and compare those results with the eigenfrequencies obtained through numerical integration of the radial differential equations. We conclude in Sec.~\ref{sec:6}. In Appendix~\ref{appendix:A}, we present the complete series expansion of the PVI $\tau$ function, reviewing the work done in \cite{Gamayun:2012ma,Gamayun:2013auu}. Finally, in Appendix~\ref{appendix:B}, we perform a comparison between the decoupled scalar-type quasinormal modes and the eigenfrequencies found by directly solving the coupled system.

\section{The Einstein-Proca system with a cosmological constant}
\label{sec:2}
A massive spin-1 field, i.e. a Proca field, minimally coupled to GR with a cosmological constant in five dimensions is described by the action
\begin{equation}\label{eq:action}
    S=\int d^5x \sqrt{-g} \left(\frac{R-2\Lambda}{16\pi}-\frac{1}{2}\mu^2A_\mu A^\mu -\frac{1}{4}F_{\mu \nu}F^{\mu \nu}\right) \,,
\end{equation}
where $g$ is the determinant of the metric $g_{\mu \nu}$, $R=R_{\mu \nu}g^{\mu \nu}$ is the Ricci scalar, with $R_{\mu \nu}$ being the Ricci tensor, $\Lambda$ is the cosmological constant, defined in terms of the characteristic AdS length, $L$, as $\Lambda=-\frac{6}{L^2}$, $A_\mu$ is the Proca field with mass $\mu$ and $F_{\mu \nu} \equiv \nabla_{\mu}A_\nu-\nabla_{\nu}A_\mu$ is
the Proca field strength, where $\nabla_\mu$ denotes the covariant derivative with respect to $x^\mu$. The Einstein field equations associated to the action Eq.~\eqref{eq:action} are 
\begin{equation}\label{eq:efe}
    G_{\mu \nu} - \frac{6}{L^2}g_{\mu \nu} = 8 \pi T_{\mu \nu}\,,
\end{equation}
where $G_{\mu \nu} = R_{\mu \nu} - \frac{1}{2} g_{\mu \nu} R$ is the Einstein tensor, and $T_{\mu \nu}$ is the stress-energy tensor, given by
\begin{equation}
    T_{\mu \nu} = g^{\alpha \beta} F_{\mu \alpha} F_{\nu \beta} +\mu^2 A_\mu A_\nu - g_{\mu \nu}\left(\frac{1}{4}F_{\alpha\beta}F^{\alpha\beta}+\frac{\mu^2}{2}A_\alpha A^\alpha\right)\,.
\end{equation}
In turn, the Proca field equations are 
\begin{equation}\label{eq:proca_equation}
    \nabla_\nu F^{\mu \nu} + \mu^2 A^\mu = 0\,.
\end{equation}
It follows from Eq.~\eqref{eq:proca_equation} that the Bianchi identity $\nabla_\mu A^\mu = 0$ is satisfied whenever $\mu \neq 0$, i.e. for a Proca field, in which case $A_\mu$ has four physical degrees of freedom. In contrast, when $\mu=0$, i.e. for a Maxwell field, the field equations are invariant under a gauge transformation, and one of the previous degrees of freedom becomes non-physical.

We consider small perturbations in the Proca field and solve Eqs.~\eqref{eq:efe} and~\eqref{eq:proca_equation} up to first order in $A_\mu$. Since the stress-energy tensor $T_{\mu \nu}$ is quadratic in $A_\mu$, Eq.~\eqref{eq:efe} completely decouples from the Proca equations Eq.~\eqref{eq:proca_equation} and reduces to the Einstein field equations with cosmological constant in vacuum. In essence, this means that the Proca perturbation does not backreact on the metric. The metric $g_{\mu \nu}$ thus remains unperturbed and corresponds to the background metric, whereas Eq.~\eqref{eq:proca_equation} is to be solved in this fixed background for $A_\mu$. In what follows, this fixed background is taken as the Schwarzschild-AdS$_5$ solution.

\section{Proca field perturbations in Schwarzschild-AdS$_5$}
\label{sec:3}

\subsection{Separation of the Proca equations}
\label{sec:3.1}

The line element of the Schwarzschild-AdS$_5$ black hole in Schwarzschild coordinates $x^\mu = (t,r,\theta^1,\theta^2,\theta^3)$ is
\begin{align}
	&ds^2 = -f(r)dt^2+f^{-1}(r)dr^2+r^2 d\Omega_{3}^2 \,\,,
	\label{eq:background}\\
    &f(r) = 1 + \frac{r^{2}}{L^{2}} - \left(1 + \frac{r_{h}^{2}}{L^{2}}\right)\left(\frac{r_{h}}{r}\right)^{2} = \frac{\left(r^{2} - r_{h}^{2}\right)\left(r^{2} - r_{c}^{2}\right)}{L^{2}r^{2}},
\end{align} 
where $t$ is the time coordinate, $r$ is the radial coordinate, $d\Omega_{3}^2$ is the line element of the unit $3$-sphere in polar coordinates ($\theta^1$, $\theta^2$, $\theta^3$), $f(r)$ is the blackening factor defined above, 
$r_h$ is the event horizon radius and the only positive root of $f(r)$, 
$r_c = i \sqrt{r_h^2 + L^2}$ is one of the imaginary roots of $f(r)$.

In spherically symmetric spacetimes, the Proca field equations reduce to a set of radial wave-like equations. This is achieved by decomposing the Proca field according to its tensorial behaviour on the sphere. We adopt the following ansatz for the Proca field $A_\mu$ 
\begin{align}\label{eq:procaansatz}
		A_\mu dx^\mu =& \,r^{-\frac{3}{2}}\sum_{\vec{k}_s} 
		\biggl(u_{0 \vec{k}_s}(t,r)dt
		+  \frac{u_{1\vec{k}_s}(t,r)}{f(r)}dr\biggr)Y_{\vec{k}_s}
        +  r^{-\frac{3}{2}}\sum_{\vec{k}_s}
		\biggl[\frac{r u_{2\vec{k}_s}(t,r)}{\ell(\ell+2)} 
		\hat{\nabla}_i Y_{\vec{k}_s} 
		d\theta^i\biggr]\nonumber \\
		& + r^{-\frac{1}{2}}\sum_{\vec{k}_v} u_{3\vec{k}_v}(t,r) Y_{\vec
		{k}_v\,i} d\theta^i \,, 
\end{align}
where $u_{0\vec{k}_s}$, $u_{1\vec{k}_s}$, $u_{2\vec{k}_s}$ and 
$u_{3\vec{k}_v}$ are functions of $t$ and $r$, 
$Y_{\vec{k}_s}$ are the scalar spherical harmonics,
$Y_{\vec{k}_v\,i}$ are the vector spherical harmonics and,
$\vec{k}_s$ and $\vec{k}_v$ are the vectors with the angular momentum 
numbers of the scalar and vector harmonics. We shall suppress the vectors $\vec{k}_s$ and $\vec{k}_v$ from now on, i.e. the functions $u_{0\vec{k}_s}$, $u_{1\vec{k}_s}$, $u_{2\vec{k}_s}$ and 
$u_{3\vec{k}_v}$ are relabelled as 
$u_{0}$, $u_{1}$, $u_{2}$ and 
$u_{3}$. Using the ansatz Eq.~\eqref{eq:procaansatz} in the Proca equations, 
we get four equations. There are three equations in the scalar-type sector
\begin{align}
    &\mathcal{D}_{\ell} u_0 +\frac{f}{r^2}\left(1-f+\frac{r}{2}\frac{df}{dr}\right)u_0
    +\frac{df}{dr}\left(\partial_t u_1-\partial_{r_*} u_0\right) = 0 \,,\label{eq:u0}\\
    &\mathcal{D}_{\ell} u_1 + f\left(\frac{1}{r^2}-\frac{4}{r^2}f+\frac{2}{r}
	\frac{df}{dr}\right)u_1
    -\frac{f}{r}\left(\frac{df}{dr}-\frac{2f}{r}\right)u_2 = 0 \,,\label{eq:u1}\\
    &\mathcal{D}_{\ell} u_2 + \frac{f}{r^2} u_2+\frac{2f\ell(\ell+2)}{r^2}u_1 = 0 \,,
	\label{eq:u2}
\end{align}
where $\mathcal{D}_\ell$ is the operator defined by
\begin{align}\label{eq:Doperator}
	\mathcal{D}_\ell= &-\partial^2_t +f^2\partial^2_{r} + f \frac{df}{dr}\partial_r-f
		\biggl(\frac{(\ell+1)^2}{r^2} - \frac{1}{4r^2}f
		+\frac{1}{2r}\frac{df}{dr} + \mu^2 \biggr)\,,
\end{align}
and the Bianchi identity of the Proca field can be used to relate the function $u_0$ with the functions $u_1$ and $u_2$ as
\begin{align}
\partial_t u_0- f\partial_{r} u_1 = \frac{f}{r}\left(\frac{3}{2}u_1-u_2\right) \,.
\label{eq:u_bianchi}
\end{align}
On the vector-type sector, there is the equation
\begin{equation}\label{eq:uv}
    \mathcal{D}_\ell u_3=0\quad.
\end{equation}
As we are interested in dynamic solutions to the Proca equations, the 
Bianchi identity in Eq.~\eqref{eq:u_bianchi} can be used to determine 
$u_0$ in terms of $u_1$, $u_2$ and $u_3$. Therefore, the scalar-type sector can 
be described solely by $u_1$ and $u_2$ that satisfy Eqs.~\eqref{eq:u1} and~\eqref{eq:u2}, while the vector-type sector can be described by 
$u_3$ that satisfies Eq.~\eqref{eq:uv}.


\subsection{The scalar-type sector: decoupling and radial equation}
\label{sec:scalardecoupeqs}

The scalar-type sector of the Proca field is described by the functions
$u_1(t,r)$ and $u_2(t,r)$. Since we 
are interested in the quasinormal modes, we assume the ansatz 
$u_1(t,r) = \mathrm{e}^{-i \omega t} u_1(r)$ and $u_2(t,r) = 
\mathrm{e}^{-i\omega t}u_2(r)$, which is equivalent to a Fourier transformation.
The system of equations, Eqs.~\eqref{eq:u1} and~\eqref{eq:u2}, becomes 
\begin{align}
    & f^{2}\partial_{r}^{2} u_{1} + f\,\frac{df}{dr}\partial_{r} u_{1} + (\omega^2 - V_{s11})u_1 - V_{s12}u_2 = 0\,\,,\label{eq:u1fourier}\\ 
    & f^{2}\partial_{r}^{2} u_{2} + f\,\frac{df}{dr}\partial_{r} u_{2} + (\omega^2 - V_{s22})u_2 - V_{s21}u_1 = 0\,\,,\label{eq:u2fourier}
\end{align}
where the potentials are given by
\begin{align}
    V_{s 1 1}=&f
		\left(\frac{3+4\mu^2 L^2}{4L^2}+ \frac{4\ell(\ell+2)+15}{4r^2}
        -\frac{27}{4r_h^2}\left(1+\frac{r_h^2}{L^2}\right)
		\left(\frac{r_h}{r}\right)^{4}\right)\,,\nonumber\\
	V_{s 1 2}=&f
	\left(-\frac{2}{r^2}+\frac{1}{r_h^2}\left(1+\frac{r_h^2}{L^2}\right)
	\left(\frac{r_h}{r}\right)^{4}\right)\,,\nonumber\\
	V_{s 2 1}=&-f\frac{2\ell(\ell+2)}{r^2}\,,\nonumber\\
	V_{s 2 2}=&f
		\left(\frac{3+4\mu^2 L^2}{4L^2}+\frac{4\ell(\ell+2)-1}{4r^2}
		-\frac{5}{4r_h^2}\left(1+\frac{r_h^2}{L^2}\right)\left(\frac{r_h}{r}
		\right)^{4}\right)\,.\label{eq:matrixpotentialu1u2}
\end{align}
The above system is coupled 
in a non-trivial way for the functions $u_1(r)$ and $u_2(r)$. It turns out that the 
system can be decoupled, motivated by the FKKS ansatz in the Schwarzschild-AdS limit.
The ansatz is able to separate the Proca equations in general rotating spacetimes in 
higher dimensions using the existence of a conformal Killing-Yano two form $h_{ab}$ and 
the ansatz for the Proca field is $A^a = B^{ab}\nabla_b Z$, where $B^{ab}$ is the 
polarization tensor satisfying $B^{ab}(g_{bc} - \beta h_{bc}) = \delta^a_c$, $\beta$ 
is the complex polarization parameter, and $Z$ is a scalar given by $Z=R(r)S(\theta_1,\theta_2,\theta_3) 
\mathrm{e}^{-i\omega t}$, for a radial function $R(r)$ and an 
angular function $S(\theta_1,\theta_2,\theta_3)$. This Proca ansatz separates the Proca equations 
into an equation for $R(r)$, an equation for $S(\theta_1,\theta_2,\theta_3)$ and a 
relation between separation constants and $\beta$. By performing the Schwarzschild limit 
directly and only looking into the scalar-type sector, the separation constants will be determined 
by the fact that the angular function becomes a spherical harmonic, due to the 
spherical symmetry imposed. The polarization is then obtained in terms of the eigenvalues of the 
spherical harmonics and the mass of the Proca field. Another way to obtain the decoupling is by 
assuming the ansatz $A^a = B^{ab}\nabla_b Z$ for the scalar-type modes, which in the Schwarzschild-limit translates into
\begin{subequations}\label{eq:lin_transf}
\begin{align}
    \begin{split}
    u_1 &= \frac{f r^{\frac{3}{2}} \partial_r R}{1 - \beta^2 r^2} - \frac{i \omega \beta r^{\frac{5}{2}}}{1-\beta^2 r^2}R\,, \label{eq:transformu1}
    \end{split}\\
    \begin{split}
    u_2 &= \ell (\ell + 2)r^{\frac{1}{2}}R\,,\label{eq:transformu2}
    \end{split}
\end{align}
\end{subequations}
which gives already a transformation for the decoupling of the scalar-type modes. 
Indeed, putting the ansatz in Eqs.~\eqref{eq:transformu1} and~\eqref{eq:transformu2} into 
Eq.~\eqref{eq:u2fourier}, one obtains the equation for $R$ as
\begin{align}\label{eq:equationR}
    \partial_r\left( \frac{f r^{3}}{1-\beta^2 r^2}\partial_r R\right)
    + \frac{r^{3}}{1-\beta^2 r^2}\left(\frac{\omega^2}{f} - \frac{\ell (\ell +2)}{r^2}
    - \mu^2\right)R - \frac{2 i \omega \beta r^{3}}{(1 - \beta^2 r^2)^2}R = 0\,\,.
\end{align}
Equation~\eqref{eq:u1fourier} must also be satisfied for the ansatz in Eqs.~\eqref{eq:transformu1} and~\eqref{eq:transformu2}. By using the equation for 
$R$ in Eq.~\eqref{eq:equationR}, Eq.~\eqref{eq:u1fourier} is only satisfied for the following 
values of $\beta$  
\begin{align}
    \beta_\pm = i\frac{2\omega \pm \sqrt{4 \omega^2 + 4 \mu^2 \ell(\ell + 2)}}
    {2\ell(\ell + 2)} \,\,,\label{eq:betapm}
\end{align}
where $\beta_{+}$ and $\beta_{-}$ select the
``electromagnetic'' and ``non-electromagnetic'' 
polarizations of the scalar-type sector, respectively. The ``non-electromagnetic'' polarization can be seen by the fact that setting $\mu=0$ yields $\beta_{-}=0$ and correspondingly the equation for $R$ in Eq.~\eqref{eq:equationR} reduces to the scalar field equation. 
Moreover, the ``electromagnetic'' polarization is defined for $\ell\geq 1$ and the ``non-electromagnetic'' polarization is defined for $\ell\geq0$. 
The monopole mode, i.e., $\ell=0$ mode, of the ``non-electromagnetic'' polarization must be taken with care. By performing the limit $\ell=0$ to $\beta_{-}$ in Eq.~\eqref{eq:betapm}, 
we have $\beta_{-} = - i\frac{\mu^2}{2\omega}$.

For convenience, we define the following dimensionless radial coordinate and parameters
\begin{equation}\label{eq:tilde_vars}
    \tilde{r} = \frac{r}{L}, \quad \tilde{r}_{h} = \frac{r_{h}}{L}, \quad \tilde{\omega} = L\,\omega, \quad \tilde{\mu} = L\,\mu, \quad \tilde{\beta} 
    = L\beta\,,
\end{equation}
which remove the AdS radius $L$ from the second-order ODE. In tilde variables, Eq.~\eqref{eq:equationR} takes the following form
\begin{equation}\label{eq:scalar_tilde}
    \frac{d}{d\tilde{r}}\left(\frac{\tilde{r}^{3}f(\tilde{r})}{1 - \tilde{\beta}^{2}\tilde{r}^{2}}\frac{d R}{d\tilde{r}}\right) + \frac{\tilde{r}^{3}}{1 - \tilde{\beta}^{2}\tilde{r}^{2}}\left( \frac{\tilde{\omega}^{2}}{f(\tilde{r})} - \frac{\ell(\ell + 2)}{\tilde{r}^{2}} - \tilde{\mu}^{2}\right)R - \frac{2 i \tilde{\omega}\tilde{\beta}\tilde{r}^{3}}{(1 - \tilde{\beta}^{2}\tilde{r}^{2})^{2}}R = 0\,,
\end{equation}
where $f(\tilde{r}) = \frac{\left(\tilde{r}^{2} - \tilde{r}_{h}^{2}\right)\left(\tilde{r}^{2} + 1 + \tilde{r}_{h}^{2}\right)}{\tilde{r}^{2}}$. The resulting Eq.~\eqref{eq:scalar_tilde}
possesses five regular singular points in the $\tilde{r}^{2}$ variable, located at $\tilde{r}^{2}_k \in \lbrace  0, \frac{1}{\tilde{\beta}^2},\tilde{r}_h^2, \tilde{r}_c^2, \infty \rbrace$.

One can write the radial differential equation, Eq. \eqref{eq:scalar_tilde}, 
on the Riemann sphere, by introducing a M\"{o}bius transformation
\begin{equation}\label{eq:Mobius}
\begin{split}
    z &= \frac{\tilde{r}^{2}}{\tilde{r}^{2} - \tilde{r}_{c}^{2}}= \frac{\tilde{r}^{2}}{\tilde{r}^{2} + 1 + \tilde{r}_{h}^{2}}\,,
\end{split}
\end{equation}
which maps the singular points to
\begin{subequations}
\begin{equation}\label{eq:mapsing}
    \tilde{r}_k^2\in\left\lbrace -1-\tilde{r}^{2}_{h}, 0, \frac{1}{\tilde{\beta}^{2}}, \tilde{r}_{h}^{2}, \infty \right\rbrace \mapsto z_k\in\left\lbrace \infty, 0, z_{1}, z_{2}, 1\right\rbrace\,\,,
\end{equation}
where the kth singularity in the left side corresponds to the kth singularity in the right side, $z_k$ being are the singularities in the $z$ variable and
\begin{equation}\label{eq:zees}
    z_{1} = \frac{1}{1 + \tilde{\beta}^{2}(1 + \tilde{r}_{h}^{2})}, \qquad z_{2} = \frac{\tilde{r}_{h}^{2}}{1 + 2\tilde{r}_{h}^{2}}.
\end{equation}
\end{subequations}
Near each singularity, one can expand the 
solution $R$ written in terms of the variable $z$ up to leading order as 
$R = a(z-z_k)^{\rho_{z_k}^-} + b(z-z_k)^{\rho_{z_k}^+}$, 
where $\rho_{z_k}^-$ and $\rho_{z_k}^+$ are the characteristic exponents at each singularity, with 
$a$ and $b$ being constants.
The characteristic exponents of the Frobenius solutions near to each singularity are
\begin{align}\label{eq:indicial_scalar}
&\rho^\pm_0 = 0\,\,,\,\,\rho^-_{z_1} = 0 \,\,,\,\, \rho^+_{z_1} = 2\,\,,\notag\\
&\rho^\pm_{z_2} = \pm \frac{\theta_h}{2}\,\,,\,\,\rho^\pm_1 = \frac{1}{2}
\left(1\pm \sqrt{1 + \tilde{\mu}^2}\right) \,\,,\,\, \rho^\pm_\infty = \pm \frac{\theta_c}{2}\,\,,
\end{align}
where
%
\begin{equation}\label{eq:thetas}
\theta_{c} = \frac{\tilde{\omega}\sqrt{1+\tilde{r}_{h}^{2}}}{\left(1 + 2\tilde{r}_{h}^{2}\right)}, \qquad \theta_{h} = \frac{i\,\tilde{r}_{h}\tilde{\omega}}{1 + 2\tilde{r}_{h}^{2}}\,. 
\end{equation}
Now, using an s-homotopic transformation
\begin{equation}
    R(z) = (z_2-z)^{\rho^{-}_{z_2}}(1 - z)^{\rho^{+}_{1}}y_s(z)\,\,,
\end{equation}
where $y_s(z)$ is an analytic function,
one obtains the resulting equation for $y_s(z)$ of the Heun-like form as
\begin{subequations}
\begin{equation}\label{eq:heun_scalar}
    \begin{split}
        \frac{d^{2}y_s}{dz^{2}} + \biggl[\frac{1}{z} - \frac{1}{z - z_{1}} &+ \frac{1 - \theta_{h}}{z - z_{2}} + \frac{1 + \sqrt{1 + \tilde{\mu}^{2}}}{z - 1}\biggr]\frac{dy_s}{dz}\\ 
        &+ \left(\frac{\kappa_{1}\kappa_{2}}{z(z - 1)} + \frac{z_{1}(z_{1} - 1)K_{1}}{z(z - z_{1})(z - 1)} - \frac{z_{2}(z_{2} - 1)K_{2}}{z(z - z_{2})(z - 1)}\right)y_s(z) = 0\,,
    \end{split}
\end{equation}
where
\begin{equation}
    \kappa_{1} = \frac{1}{2}\left(\theta_{h} - 1 - \sqrt{1 + \tilde{\mu}^{2}} - \theta_{c}\right), \qquad \kappa_{2} = \frac{1}{2}\left(\theta_{h} - 1 - \sqrt{1 + \tilde{\mu}^{2}} + \theta_{c}\right),
\end{equation}
\begin{equation}\label{eq:accessory_scalar_K1}
    K_{1} = - \frac{(1 + \sqrt{1 + \tilde{\mu}^{2}})}{2(z_{1} - 1)} + \frac{\theta_{h}}{2(z_{2} - z_{1})} + \frac{i (1 + \tilde{\beta}^{2}(1 + \tilde{r}_{h}^{2}))\tilde{\omega}}{2\tilde{\beta}(1 + \tilde{r}_{h}^{2})(1-\tilde{\beta}^{2}\tilde{r}_{h}^{2})}\,\,,
\end{equation}
\begin{equation}\label{eq:accessory_scalar_K2}
    \begin{split}
        K_{2} = &-\frac{\ell(\ell + 2) + \tilde{\mu}^{2}\tilde{r}_{h}^{2}}{4(1 + 2\tilde{r}_{h}^{2})z_{2}(z_{2} - 1)} - \frac{\tilde{\omega}^{2}}{4(1 + \tilde{r}_{h}^{2})} - \frac{(1 + \sqrt{1 + \tilde{\mu}^{2}})(1 - \theta_{h})}{2(z_{2} - 1)} + \frac{\theta_{h}}{2z_{2}}\\
        &-\frac{\theta_{h}}{2(z_{2} - z_{1})} + \frac{i\tilde{\beta}(1 + 2\tilde{r}_{h}^{2})\tilde{\omega}}{2(1 + \tilde{r}_{h}^{2})(1 - \tilde{\beta}^{2}\tilde{r}_{h}^{2})}\,.
    \end{split}
\end{equation}
\end{subequations}

Equation \eqref{eq:heun_scalar} describes two different radial systems, depending on each polarization $\tilde{\beta}_{\pm}$, corresponding to the two physical degrees of freedom associated with the scalar-type sector.

\subsection{The vector-type sector: radial equation}
\label{sec:vectoreqs}

The vector-type sector of the Proca field is described by the function
$u_3(t,r)$. For the treatment of the quasinormal modes, we assume the ansatz $u_3(t,r) = \mathrm{e}^{-i \omega t} u_3(r)$. The equation for $u_3(r)$ is 
\begin{align}
    & f^{2}\partial_{r}^{2} u_{3} + f\,\frac{df}{dr}\partial_{r} u_{3} + (\omega^2 - V_{v})u_3 = 0\,\,,\label{eq:u3fourier}
\end{align}
where the potential $V_v$ is given by
\begin{align}
    V_{v}=f
     \biggl(&\frac{3+4\mu^2 L^2}{4 L^2} 
     +\frac{(2\ell+1)(2\ell+3)}{4 r^2} +\frac{5}{4r_h^2}\left(1+\frac{r_h^2}{L^2}\right)\left(\frac{r_h}{r}\right)^{4}
	 \biggr)\,\,.\label{eq:potentialu3}
\end{align}
In terms of the tilde notation, the radial equation Eq.~\eqref{eq:u3fourier} 
takes the following form
\begin{equation}\label{eq:vector_tilde}
    \frac{d^{2}u_{3}}{d \tilde{r}^{2}} + \frac{1}{f(\tilde{r})}\frac{df}{d\tilde{r}}\frac{d u_{3}}{d \tilde{r}} + \left(\frac{\tilde{\omega}^{2}}{f(\tilde{r})^{2}} - \frac{\left(2\ell + 3\right)\left(2\ell + 1\right)}{4\tilde{r}^{2}f(\tilde{r})} + \frac{f(\tilde{r}) - 1}{4\tilde{r}^{2}f(\tilde{r})} - \frac{1}{2\tilde{r}f(\tilde{r})}\frac{df}{d\tilde{r}} - \frac{\tilde{\mu}^{2}}{f(\tilde{r})}\right) u_{3} \hskip-0.1cm= 0\,,
\end{equation}
where $f(\tilde{r}) = \frac{\left(\tilde{r}^{2} - \tilde{r}_{h}^{2}\right)\left(\tilde{r}^{2} + 1 + \tilde{r}_{h}^{2}\right)}{\tilde{r}^{2}}$.

In contrast to the radial equation of the scalar-type sector Eq.~\eqref{eq:scalar_tilde}, the radial equation of the vector-type sector Eq.~\eqref{eq:vector_tilde} contains one singularity less in $\tilde{r}^{2}$ variable, i.e. one has four singularities 
$\tilde{r}_k^2 \in \{0, \tilde{r}_h^2, \tilde{r}_c^2, \infty\}$. One can again work with the variable $z$ through Eq.~\eqref{eq:Mobius}. The singularities $\tilde{r}_k^2$
are then mapped to the singularities $z_k$ through 
\begin{align}
\label{eq:mapsing2}
    \tilde{r}_k^2\in\left\lbrace -1-\tilde{r}^{2}_{h}, 0, \tilde{r}_{h}^{2}, \infty \right\rbrace \mapsto z_k\in\left\lbrace \infty, 0, z_{0}, 1\right\rbrace\,\,,
\end{align}
where
\begin{equation}\label{eq:zee0}
    z_{0} = \frac{\tilde{r}_{h}^{2}}{1 + 2\tilde{r}_{h}^{2}}\,.
\end{equation}
The solution $u_3$ written 
in the $z$ variable can be expanded up to leading order as $u_3 = a(z - z_k)^{\rho^-_{z_k}} + b (z - z_k)^{\rho^+_{z_k}}$, with $\rho_{z_k}^\pm$ being the characteristic 
exponents for the vector case given by 
\begin{align}\label{eq:indicial_vector}
&\rho^-_0 = \frac{1}{4}\,\,,\,\,\rho^+_0 = \frac{5}{4}\,\,,\,\,\rho^\pm_{z_0} = \pm \frac{\theta_h}{2}\,\,,\,\,\rho^\pm_1 = \frac{1}{4}
\left(1 \pm 2\sqrt{1 + \tilde{\mu}^2}\right) \,\,,\,\, \rho^\pm_\infty = \pm \frac{\theta_c}{2}\,\,,
\end{align}
%
Using now a s-homotopic transformation to $u_3$ as 
\begin{equation}
    u_3(z) = z^{\rho^{+}_{0}}(1 - z)^{\rho^{+}_{1}}(z_{0} - z)^{\rho_{z_0}^{-}}y_v(z)\,,
\end{equation}
where $y_v(z)$ is an analytic function,
one can transform Eq.~\eqref{eq:vector_tilde} into the canonical form of 
the Heun differential equation
\begin{subequations}
\begin{equation}\label{eq:heun_vector}
    \frac{d^{2}y_{v}}{d z^{2}} + \left[\frac{2}{z} + \frac{1  - \theta_{h}}{z - z_{0}} + \frac{1 + \sqrt{1 + \tilde{\mu}^{2}}}{z - 1}\right]\frac{d y_{v}}{d z} + \left(\frac{\kappa_{1}\kappa_{2}}{z(z - 1)} - \frac{z_{0}(z_{0} - 1)K_{0}}{z(z - z_{0})(z - 1)}\right)y_{v}(z) = 0
\end{equation}
where
\begin{equation}
    \kappa_{1} = \frac{1}{2}\left(3 + \sqrt{1 + \tilde{\mu}^{2}} - \theta_{h} + \theta_{c} \right), \qquad \kappa_{2} = \frac{1}{2}\left(3 + \sqrt{1 + \tilde{\mu}^{2}} - \theta_{h} - \theta_{c} \right)\,,
\end{equation}
\begin{equation}\label{eq:accessory_vector}
    K_{0} = -\frac{\ell(\ell + 2) + \tilde{\mu}^{2}\tilde{r}_{h}^{2} + 1}{4\left(1 + 2\tilde{r}_{h}^{2}\right)z_{0}\left(z_{0} - 1\right)} - \frac{\tilde{\omega}^{2}}{4\left(1 + \tilde{r}_{h}^{2}\right)} - \frac{(1 + \sqrt{1 + \tilde{\mu}^{2}})\left(1 - \theta_{h}\right)}{2\left(z_{0} - 1\right)} - \frac{\left(1 - \theta_{h}\right)}{z_{0}}\,,
\end{equation}
\end{subequations}
where $z_{0}$ and $K_{0}$ are the conformal modulus and the accessory parameter, respectively.

\subsection{Boundary conditions}
\label{sec:3.4}

The QNMs are solutions of the eigenvalue problem relative to \eqref{eq:heun_scalar} or \eqref{eq:heun_vector}, satisfying specific boundary conditions: a purely ingoing wave at the event horizon and regularity at spatial infinity. In particular, for the radial ODE of the scalar-type sector we focus on solutions with the following asymptotic behavior
\begin{equation}\label{eq:boundaryfors}
R(z)\sim
\begin{cases}
(z_{2} - z)^{-\theta_{h}/2},& z \to z_{2}, \\
\\
A_{s}\,(1 - z)^{\thalf\left(1 - \sqrt{1 + \tilde{\mu}^{2}}\right)} + B_{s}\,(1 - z)^{\thalf\left(1 + \sqrt{1 + \tilde{\mu}^{2}}\right)},& z \to 1.
\end{cases}
\end{equation}
where $A_{s}$ and $B_{s}$ are constants. For $\tilde{\mu} > 0$, at $z \to 1$ the second solution converges, whereas the first diverges, and thus, these solutions will correspond to normalizable and non-normalizable solutions, respectively\footnote{Note that $\rho^{\pm}_{\infty}$ can be expressed in terms of the scaling dimension $\Delta$ of a dual current operator living on the boundary of AdS$_{5}$, due to the relation $(\Delta - 1)(\Delta - 3) = \tilde{\mu}^{2}$.}. To ensure regularity, we set $A_{s} = 0$.

For the radial ODE of the vector-type sector, the asymptotic behavior of the radial solutions is given by
\begin{equation}\label{eq:boundaryforv}
u_{3}(z)\sim
\begin{cases}
(z_{0} - z)^{-\theta_{h}/2},& z \to z_{0}, \\
\\
A_{v}\,(1-z)^{\thalf\left(\thalf - \sqrt{1 + \tilde{\mu}^{2}}\right)} + B_{v}\,(1-z)^{\thalf\left(\thalf + \sqrt{1 + \tilde{\mu}^{2}}\right)},& z \to 1.
\end{cases}
\end{equation}
where $A_{v}$ and $B_{v}$ are constants. Similarly to the scalar-type case, we require that $A_{v} = 0$ for $\tilde{\mu} > 0$.

In the next Section, we will reformulate the boundary value problem for the radial differential equation of the scalar and vector-type sector in terms of the initial conditions of the Painlev\'{e} VI $\tau$ function.

\section{The isomonodromy method: initial conditions on the Painlev\'{e} VI system}
\label{sec:4}

\subsection{The isomonodromic deformation of the Heun equation 
and the Painlev\'{e} VI $\tau$ function}

In this section, isomonodromic deformations of the Heun equation are reviewed 
so that it can be applied to the computation of the quasinormal modes of the Proca field.

Consider a $2 \times 2$ first-order linear system with four regular singular points $0,t,1,\infty$ on the Riemann sphere $\mathbb{P}^{1}$:
\begin{equation}\label{eq:matrixODE}
\dfrac{d\Phi}{dz} = A(z)\Phi, \qquad A(z) = \frac{A_{0}}{z} + \frac{A_{t}}{z - t} + \frac{A_{1}}{z - 1} =
\begin{pmatrix}
A_{11}(z) & A_{12}(z) \\
A_{21}(z) & A_{22}(z)
\end{pmatrix},
\end{equation}
where the $2\times 2$ matrix $\Phi$ is the fundamental matrix solution, $A_{\nu}\,\left(\nu = 0,t,1 \right)$ are $2 \times 2$ residue matrices that do not depend on $z$, and the $A_{ij}(z)$ with $i,j \in \,\{1,2\}$ are the matricial components of $A(z)$. There is a freedom on the general expression of the matrices $A_\nu$ related to the choice of basis of $\Phi$. Thus, without loss of generality, we may assume
\begin{equation}\label{eq:A_infty}
    A_{\infty} = -\left(A_{0} + A_{t} + A_{1}\right) = \begin{pmatrix}
\kappa_{1} & 0 \\
0 & \kappa_{2}
\end{pmatrix}\,.
\end{equation}
The fundamental matrix solution of Eq.~\eqref{eq:matrixODE} is composed by two linearly independent vector solutions making up the columns of $\Phi$ as
\begin{equation}
\Phi(z)=
\begin{pmatrix}
y_{11}(z) & y_{12}(z) \\
y_{21}(z) & y_{22}(z)
\end{pmatrix}
\end{equation}
Each column of $\Phi(z)$ satisfies a system of coupled first-order differential equations
\begin{equation}
\frac{d}{dz}\begin{pmatrix} y_{1j} \\ y_{2j}  \end{pmatrix} = A(z)\begin{pmatrix} y_{1j} \\ y_{2j}  \end{pmatrix}, \qquad j = 1,2\,.
\end{equation}
Expanding the first column, we get
\begin{equation}
    \begin{split}
        &\frac{d}{dz}y_{11}(z) = A_{11}(z)y_{11} + A_{12}(z)y_{21}(z)\,,\\
        &\frac{d}{dz}y_{21}(z) = A_{21}(z)y_{11} + A_{22}(z)y_{21}(z)\,.
    \end{split}
\end{equation}
Then, by taking the derivative of the first equation and substituting $y^{\prime}_{21}$ from the second equation, it is straightforward to check that $y_{11}$ satisfies the following second-order differential equation
\begin{equation}\label{eq:second_order_ODE}
\frac{d^2y_{11}}{dz^2} - \left(\Tr A + \partial_z \log A_{12}\right)\frac{dy_{11}}{dz} + \left(\det A - \partial_z A_{11} + A_{11}\partial_z \log A_{12}\right)y_{11} = 0\,,
\end{equation}
and there is an analogous equation for $y_{12}$\footnote{We will drop the indices in $y$ to not overcrowd the notation.}. Following the seminal works of Jimbo et al.~\cite{Jimbo:1981tov,Jimbo:1981zz}, 
we introduce the parametrization for the residue matrices
\begin{equation}\label{eq:A_nu}
A_{\nu} =
\begin{pmatrix}
p_{\nu} + \vartheta_{\nu} & -q_{\nu}\,p_{\nu} \\
\frac{p_{\nu} + \vartheta_{\nu}}{q_{\nu}} & -p_{\nu}
\end{pmatrix}\,, 
\qquad \nu =\lbrace 0,t,1 \rbrace
\end{equation}
where $\Tr A_{\nu} = \vartheta_{\nu}$ and $\det A_{\nu} = 0$. Using $\kappa_{1} + \kappa_{2} = -(\vartheta_{0} + \vartheta_{t} + \vartheta_{1})$ and $\vartheta_{\infty} = \kappa_{1} - \kappa_{2}$, the diagonal terms of $A_{\infty}$ are 
\begin{equation}
    \kappa_{1} = \frac{1}{2}(\vartheta_{\infty} - \vartheta_{0} - \vartheta_{1} - \vartheta_{t})\,, \qquad \kappa_{2} = -\frac{1}{2}(\vartheta_{\infty} + \vartheta_{0} + \vartheta_{1} + \vartheta_{t})\,.
\end{equation}
The parameters $q_{\nu}, p_{\nu}$ are subject to extra constraints due to the choice of $A_{\infty}$ being diagonal,
\begin{equation}\label{eq:constraints}
    \sum_{\nu} p_{\nu} = \kappa_{2}\,, \qquad \sum_{\nu} q_{\nu}\,p_{\nu} = 0\,, \qquad \sum_{\nu} \frac{p_{\nu} + \vartheta_{\nu}}{q_{\nu}} = 0\,,
\end{equation}
where the second equation in Eq.~\eqref{eq:constraints} implies that the entry $A_{12}$ is of the form 
\begin{equation}\label{eq:A_{12}}
    A_{12}(z) = \frac{(A_{0})_{12}}{z} + \frac{(A_{t})_{12}}{z - t} + \frac{(A_{1})_{12}}{z - 1} = \frac{k(z - \lambda)}{z(z - t)(z - 1)}\,, \qquad k \in \mathbb{C}
\end{equation}
and $z = \lambda$ corresponds to a simple zero of $A_{12}(z)$. Furthermore, to fully solve the system for $q_{\nu}$ and $p_{\nu}$, we introduce
\begin{equation}\label{eq:mu}
    \eta \coloneqq A_{11}(z=\lambda) = \dfrac{(A_{0})_{11}}{\lambda} + \dfrac{(A_{t})_{11}}{\lambda-t} + \dfrac{(A_{1})_{11}}{\lambda-1}\,,
\end{equation}
which allows us to determine the $q_{\nu}$ and $p_{\nu}$ in terms of $\left(\lambda,\eta,t\right)$. Their explicit forms can be found in \cite{Jimbo:1981zz}. It turns out that by replacing Eqs.~\eqref{eq:A_infty}, \eqref{eq:A_{12}} and \eqref{eq:mu} into Eq.~\eqref{eq:second_order_ODE}, we obtain an equation with an extra singularity at $z = \lambda$ of the form
\begin{subequations}
\begin{equation}\label{eq:deformed_heun}
\begin{split}
    \frac{d^2 y}{dz^2} &+ \left(\dfrac{1 - \vartheta_{0}}{z} + \dfrac{1 - \vartheta_{1}}{z-1} + 
    \dfrac{1 - \vartheta_{t}}{z-t} - \dfrac{1}{z - \lambda}\right)\frac{dy}{dz} \\
    &+ \biggl\lbrace\dfrac{\kappa_{1}(\kappa_{2} + 1)}{z(z-1)} - \dfrac{t(t-1)K}{z(z-1)(z-t)} + \dfrac{\lambda(\lambda-1)\eta}{z(z-1)(z - \lambda)} \biggr\rbrace y = 0\,,
\end{split}
\end{equation}
where
\begin{equation}\label{eq:kee}
    K = H + \dfrac{\lambda(\lambda-1)}{t(t-1)}\eta + \dfrac{(\lambda-t)}{t(t-1)}\kappa_{1}\,,
\end{equation}
\begin{equation}\label{eq:h}
    H =  \dfrac{1}{t}\Tr (A_0 A_t) + \dfrac{1}{t-1}\Tr (A_1  A_t) -\dfrac{1}{t}\vartheta_{0}\vartheta_{t}  -\dfrac{1}{t-1}\vartheta_{1}\vartheta_{t}\,.
\end{equation}
\end{subequations}
and we refer to Eq.~\eqref{eq:deformed_heun} as the deformed Heun equation. The singular point at $z = \lambda$ has characteristic exponents $\{0,2\}$, 
and it is a non-logarithmic singular point if and only if
\begin{equation}\label{eq:Kamiltonian}
K(\lambda,\eta,t) = \frac{\lambda(\lambda-1)(\lambda-t)}{t(t-1)}\biggl[
\eta^2 - \left(\frac{\vartheta_{0}}{\lambda} + \frac{\vartheta_{t} - 1}{\lambda-t} + \frac{\vartheta_{1}}{\lambda-1}
\right)\eta +\frac{\kappa_{1}(\kappa_{2} + 1)}{\lambda(\lambda-1)}\biggr],
\end{equation}
and hence corresponds to an apparent singularity. Moreover, it has been shown \cite{Garnier:1912} that Eq.~\eqref{eq:Kamiltonian} defines a Hamiltonian system
\begin{equation}\label{eq:Ksystem}
\dfrac{d\lambda}{dt} = \frac{\partial K}{\partial \eta}\,, \qquad \dfrac{d\eta}{dt} = -\frac{\partial K}{\partial \lambda}\,,
\end{equation} 
where $(\lambda,\eta)$ are canonically conjugated coordinates, and the equation of motion for $\lambda(t)$ satisfies the Painlev\'{e} VI (PVI) equation:
\begin{equation}\label{eq:PainleveVI}
\begin{split}
\frac{d^2\lambda}{dt^2} = &\dfrac{1}{2}\left(\dfrac{1}{\lambda} + \dfrac{1}{\lambda-1} + \dfrac{1}{\lambda-t}\right)\left(\frac{d\lambda}{dt}\right)^{2}-\left(\dfrac{1}{t}+\dfrac{1}{t-1}+\dfrac{1}{\lambda-t}\right)\frac{d\lambda}{dt} +\\
&+ \dfrac{\lambda(\lambda-1)(\lambda-t)}{2t^{2}(t-1)^{2}}\left[\left(\vartheta_{\infty}-1\right)^{2} -\dfrac{\vartheta^{2}_{0}t}{\lambda^{2}} + \dfrac{\vartheta^{2}_{1}(t-1)}{(\lambda-1)^{2}} - \dfrac{(\vartheta^{2}_{t} - 1)t(t-1)}{(\lambda-t)^{2}}\right]\,,
\end{split}
\end{equation}
the most general nonlinear second-order differential equation that enjoys the Painlev\'{e} property: the singularities of the solutions, apart from $t=0,1,\infty$, are simple poles and depend on the initial conditions \cite{Iwasaki:2013gauss}.

Alternatively, Schlesinger \cite{Schlesinger:1912} showed that as a consequence of the isomonodromy condition, the residue matrices $A_{\nu}$ satisfy a set of non-linear differential equations, also known as Schlesinger equations
\begin{equation}\label{eq:Schlesinger}
		\frac{d A_{0}}{dt} = -\frac{\left[A_{0},A_{t}\right]}{t}, \qquad
		\frac{d A_{1}}{dt} = -\frac{\left[A_{1},A_{t}\right]}{t - 1}, \qquad
		\frac{d A_{t}}{dt} = \frac{\left[A_{0},A_{t}\right]}{t} + \frac{\left[A_{1},A_{t}\right]}{t - 1},
\end{equation}
where the last equation implies that $A_{\infty} = \mathrm{const}$. Then the entry $A_{12}(z)$ is of the form in Eq.~\eqref{eq:A_{12}}, and the Schlesinger equations in Eq.~\eqref{eq:Schlesinger} reduce to the PVI equation Eq.~\eqref{eq:PainleveVI}. This equation represents the isomonodromic deformation equation of the Fuchsian system in Eq,~\eqref{eq:matrixODE}, as it governs how the positions of singular points can change while preserving the monodromy data. Furthermore, the Hamiltonian Eq.~\eqref{eq:Kamiltonian} generates an isomonodromic flow in terms of $(\lambda(t),\eta(t))$, and admits the definition of a $\tau$ function
\begin{equation}\label{eq:tauf}
    \frac{d}{d t}\log \tau(\rho;t) = \frac{1}{t}\Tr\,A_{0}A_{t} + \frac{1}{t - 1}\Tr\,A_{t}A_{1} - \dfrac{1}{2t}\vartheta_{0}\vartheta_{t} - \dfrac{1}{2(t-1)}\vartheta_{1}\vartheta_{t}\,,
\end{equation}
where the logarithmic derivative of PVI $\tau$ function solves a nonlinear second-order ODE called $\sigma$-form of Painlev\'{e} VI, which can be found in \cite{Jimbo:1981zz}, but is beyond the scope of this work. Moreover, this isomonodromic $\tau$ function can be written in terms of monodromy data $\rho$ associated with the $2 \times 2$ Fuchsian system with four regular singular points. In fact, the monodromy group of functions on the complex plane will play a central role in understanding the deformation theory, see \cite{Iwasaki:2013gauss}.

\subsection{Connection of Painlev\'e $\tau$ function with monodromy data}

In order to connect Eq.~\eqref{eq:tauf} with the monodromy data, we must give a description of the monodromy group of the four-punctured Riemann sphere. The fundamental matrix solution $\Phi(z)$ is multivalued on $\mathbb{P}^{1} \setminus \lbrace 0,t,1,\infty \rbrace$, as its analytic continuation around a closed path $\gamma$, enclosing one or more singular points, produces non-trivial monodromy. We associate each path enclosing only one singular point $\gamma_{i}$ with a monodromy matrix $M_{i}$, and label those matrices by their trace $m_{i} = \Tr M_{i} = 2\cos(\pi\vartheta_{i})$. The monodromy group is then generated by three out of four monodromy matrices $M_{0,t,1} \in G = SL(2,\mathbb{C})$ obeying 
\begin{equation}
    M_{0}M_{1}M_{t}M_{\infty} = \mathbb{1}\,,
\end{equation}
since the composition of the monodromies over all singular points is a contractible curve. In order to fully characterize the $M_{i}$ (up to an overall conjugation), we introduce the composite monodromies $\sigma_{ij}$ as
\begin{equation}
    m_{ij} = 2\cos\pi\sigma_{ij} = \Tr M_{i}M_{j}\,, \quad i,j = 0,t,1\,,
\end{equation}
where $M_{i}M_{j}$ represents the analytic continuation around two singular points. Furthermore, the seven invariant functions $(m_{i},m_{ij})$ satisfy the Fricke-Jimbo cubic relation
\begin{equation}\label{eq:fricke-jimbo}
    \begin{split}
        m_{0t}^{2} + m_{t1}^{2} &+ m_{01}^{2} + m_{0t}m_{t1}m_{01} + m_{0}^{2} + m_{t}^{2} + m_{1}^{2} + m_{\infty}^{2} + m_{0}m_{t}m_{1}m_{\infty}\\
        &= (m_{0}m_{t} + m_{1}m_{\infty})m_{0t} + (m_{1}m_{t} + m_{0}m_{\infty})m_{t1} + (m_{0}m_{1} + m_{t}m_{\infty})m_{01} + 4\,,
    \end{split}
\end{equation}
which fixes one of the composite monodromies, say $m_{01}$, given the other two, $m_{0t}$ and $m_{t1}$. Hence, one can define the monodromy data of the four-punctured Riemann sphere as $\rho = \lbrace \vartheta,\sigma \rbrace = \lbrace \vartheta_{0}, \vartheta_{t},\vartheta_{1},\vartheta_{\infty},\sigma_{0t}=\sigma,\sigma_{t1}\rbrace$. Given local monodromies, solutions of the Painlev\'{e} VI equation can be parameterized by any pair of parameters $\sigma_{ij}$, for example, $\lambda(t) = \lambda(t,\sigma,\sigma_{t1})$ \cite{Jimbo:1982,Guzzetti:2012}. 

In fact, one can relate a solution of the isomonodromic flow $(\lambda(t),\eta(t))$ to the PVI $\tau$ function by replacing Eqs.~\eqref{eq:tauf} and \eqref{eq:kee} into Eq.~\eqref{eq:Kamiltonian}, and then taking its derivative. The resulting conditions are a set of transcendental equations for the two integration constants $\sigma$ and $\sigma_{t1}$
\begin{subequations}\label{eq:conditions}
\begin{equation}\label{eq:first_cond}
    \begin{split}
        \frac{d}{d t}\log \tau(\rho;t)  = &\frac{\lambda(\lambda-1)(\lambda-t)}{t(t-1)}\biggl[\eta^2 - \left(\frac{\vartheta_{0}}{\lambda} + \frac{\vartheta_{t}}{\lambda-t} + \frac{\vartheta_{1}}{\lambda-1} \right)\eta +\frac{\kappa_{1}\kappa_{2}}{\lambda(\lambda-1)}\biggr]\\
        &+ \dfrac{1}{2t}\vartheta_{0}\vartheta_{t} + \dfrac{1}{2(t-1)}\vartheta_{1}\vartheta_{t}\,,
    \end{split}
\end{equation}
\begin{equation}\label{eq:second_cond}
    \begin{split}
        \frac{d}{dt}\left[t(t-1)\frac{d}{d t}\log \tau(\rho;t)\right]  = &-\frac{\lambda(\lambda-1)(\lambda-t)^{2}}{t(t-1)}\biggl[\eta^2 - \left(\frac{\vartheta_{0}}{\lambda} + \frac{\vartheta_{t} - \vartheta_{\infty}}{\lambda-t} + \frac{\vartheta_{1}}{\lambda-1} \right)\eta \\&+\frac{\kappa_{1}^{2}}{(\lambda-t)^{2}}\biggr]
        -\frac{\lambda-1}{t-1}\kappa_{1}\vartheta_{0} - \frac{\lambda}{t}\kappa_{1}\vartheta_{1} - \kappa_{1}\kappa_{2} + \dfrac{1}{2}(\vartheta_{0} + \vartheta_{1})\vartheta_{t}\,.
    \end{split}
\end{equation}
\end{subequations}

\subsection{Quasinormal modes using the isomonodromic flow}

For the computation of the quasinormal modes
at hand, the isomonodromic flow $(\lambda(t),\eta(t))$ is considered and 
initial conditions are imposed. Among these initial conditions, the 
parameter $t$ is associated to the singularity correspondent to the 
event horizon. For small horizon radius, the parameter $t$ starts 
very small and through the isomonodromic flow decreases towards $t=0$. 
Since $\lambda(t)$ is an apparent singularity, 
the Fuchsian system at the limit of $t=0$ 
can be solved in terms of hypergeometric functions~\cite{Jimbo:1982,Its:2016jkt}.
One can then find a series expansion of the PVI $\tau$ function 
valid in a neighbourhood of $t=0$, which is found by using the properties 
of the hypergeometric functions and
solves the $\sigma$-form of the PVI equation, see App.~\ref{appendix:A}.
The local monodromies that parametrize the expansion of the PVI $\tau$ function 
are then 
replaced by the characteristic exponents of the radial differential equations 
describing the perturbations of the Proca field. 
Finally, the boundary conditions of the radial systems
can be encoded into a composite monodromy matrix around two singular points $r = r_{h}$ and $r = \infty$ of triangular form \cite{BarraganAmado:2018zpa}. The boundary conditions then impose 
\begin{equation}
    \sigma_{t1} = \vartheta_{t} + \vartheta_{1} + 2n\,, \quad n \in \mathbb{Z}\,,   
\end{equation}
for the composite monodromy $\sigma_{t1}$, which can be thought of as a radial quantization condition. Having the PVI $\tau$ function, 
one has two quantities, 
$\vartheta_t,\vartheta_{\infty}$ containing the quasinormal mode frequency and the composite monodromy 
$\sigma$, that must be found by solving the nonlinear system in Eqs.~\eqref{eq:conditions}. 
This summarizes the reduction of the boundary value problem for the differential 
equations to the resolution of a nonlinear system of algebraic equations.


In what follows, we will elaborate on the initial conditions for the isomonodromic flow in the case of the scalar-type and vector-type radial ODE. Namely, we will be interested in the deformed Heun equation as a consequence of introducing the separation parameter $\beta$. As it was elucidated in the case of Maxwell perturbations on Kerr-AdS$_{5}$ black holes \cite{Amado:2020zsr}, there is a non-trivial relation between the apparent singularity of the Fuchsian system $\lambda$ and the parameter $\beta$ in the radial ODE.

\subsection{The radial equation of the scalar-type sector as a deformed Heun equation}
\label{sec:4.1}
One can recover the radial ODE of the scalar-type sector in Eq.~\eqref{eq:heun_scalar} from the deformed Heun equation 
\begin{equation}\label{eq:deformed}
    \begin{split}
        \frac{d^2y}{dz^2} &+ \left(\dfrac{1 - \vartheta_{0}}{z} + \dfrac{1 - \vartheta_{1}}{z-1} + \dfrac{1 - \vartheta_{t}}{z-t} - \dfrac{1}{z - \lambda}\right)\frac{dy}{dz} \\
        &+ \biggl\lbrace\dfrac{\kappa_{1}(\kappa_{2} + 1)}{z(z-1)} - \dfrac{t(t-1)K}{z(z-1)(z-t)} + \dfrac{\lambda(\lambda-1)\eta}{z(z-1)(z - \lambda)} \biggr\rbrace y = 0\,,
    \end{split}
\end{equation}
by 
setting the initial conditions to the Hamiltonian system in Eq.~\eqref{eq:Ksystem} as
\begin{equation}\label{eq:initial_scalar}
    t = z_{2}\,, \qquad \lambda(t = z_{2}) = z_{1}\,, \qquad \eta(t = z_{2}) = K_{1}\,,
\end{equation}
with $y(t=z_2;z) = y_s(z)$,
and setting the local isomonodromies as 
\begin{align}\label{eq:monoinit}
    \vartheta_0 = 0\,\,,\,\,\vartheta_t = \theta_h\,\,,\,\,\vartheta_1 = -\sqrt{1+\mu^2}\,\,,
    \,\,\vartheta_\infty = \theta_c +1\,\,,
\end{align}
where $\theta_{h}$, $\theta_{c}$, $z_{1}$, $z_{2}$, $K_{1}$, and $K_{2}$ are given by Eqs.~\eqref{eq:thetas}, \eqref{eq:zees}, \eqref{eq:accessory_scalar_K1}, and \eqref{eq:accessory_scalar_K2}, respectively. As a check of consistency, 
with the initialization in Eq.~\eqref{eq:initial_scalar} and the monodromies in 
Eq.~\eqref{eq:monoinit}, one obtains $K(t=z_2) = K_2$ as it should from Eq.~\eqref{eq:heun_scalar}.
The expansion of the PVI $\tau$ function in App.~\ref{appendix:A}, 
for fixed $\tilde{r}_h$, $\tilde{\mu}$ and $\ell$, 
is completely determined up to the local isomonodromies $\vartheta_t,\vartheta_{\infty}$, which depend on the 
quasinormal frequency, and the composite monodromy $\sigma$. These two quantities can then 
be computed using the initial conditions for the PVI $\tau$ function Eq.~\eqref{eq:conditions}, 
which for this case can be written as
\begin{subequations}\label{eq:conditions_scalar}
\begin{equation}
    \begin{split}
        \frac{d}{d t}\log \tau(\rho;t)\biggr\vert_{t = z_{2}}  = &\frac{z_{1}(z_{1} - 1)(z_{1} - z_{2})}{z_{2}(z_{2} - 1)}\biggl[K_{1}^{2} - \biggl(\frac{\theta_{h}}{z_{1} - z_{2}} -\frac{\sqrt{1 + \tilde{\mu}^{2}}}{z_{1} - 1}\biggr)K_{1}\\
        &+\frac{\frac{1}{4}((\theta_{h} - 1 - \sqrt{1 + \tilde{\mu}^{2}})^{2} - \theta_{c}^{2})}{z_{1}(z_{1} - 1)}\biggr] - \dfrac{\sqrt{1 + \tilde{\mu}^{2}}}{2(z_{2} - 1)}\theta_{h}\,,
    \end{split}
\end{equation}
\begin{equation}
    \begin{gathered}
        \frac{d}{dt}\biggl[t(t-1)\frac{d}{d t}\log \tau(\rho;t)\biggr]\biggr\vert_{t = z_{2}} \hskip-0.5cm = -\frac{z_{1}(z_{1} - 1)(z_{1} - z_{2})^{2}}{z_{2}(z_{2} - 1)}\biggl[K_{1}^2 - \left(\frac{\theta_{h} - \theta_{c} - 1}{z_{1} - z_{2}} - \frac{\sqrt{1 + \tilde{\mu}^{2}}}{z_{1} - 1} \right)K_{1}\\
        +\frac{\frac{1}{4}(\theta_{h} - 1 - \sqrt{1 + \tilde{\mu}^{2}} - \theta_{c})^{2}}{(z_{1} - z_{2})^{2}}\biggr] + \frac{z_{1}}{z_{2}}\frac{\sqrt{1 + \tilde{\mu}^{2}}}{2}(\theta_{h} - 1 - \sqrt{1 + \tilde{\mu}^{2}} - \theta_{c})\\ 
        - \frac{1}{4}((\theta_{h} - 1 - \sqrt{1 + \tilde{\mu}^{2}})^{2} - \theta_{c}^{2}) -\frac{\sqrt{1 + \tilde{\mu}^{2}}}{2}\theta_{h}\,,
    \end{gathered}
\end{equation}
\end{subequations}
As a result, Eqs.~\eqref{eq:conditions_scalar} are fully determined by the physical parameters $(\tilde{\omega},\ell,\tilde{\mu},\tilde{r}_{h})$, for the corresponding $\tilde{\beta}_{\pm}$ mode. One then has to solve Eqs.~\eqref{eq:conditions_scalar} 
to obtain the pair $(\tilde{\omega},\sigma)$. One must notice that one requires 
two equations instead of just one since the composite monodromy $\sigma$ is not 
determined apriori.


\subsection{The radial equation of the vector-type sector as a Heun equation}
\label{sec:4.2}
By inspection of Eq.~\eqref{eq:deformed}, we note that the coalescence of the apparent singularity at $z =\lambda$ with one of the singular points $z=\lbrace 0,t,1,\infty\rbrace$ reduces the deformed Heun equation to a Heun equation. 
The radial ODE of the vector-type sector in Eq.~\eqref{eq:heun_vector} for $y_v(z)$ 
can then be given by 
Eq.~\eqref{eq:deformed} for $y(t,z)$ 
with a special set of initial conditions to the isomonodromic flow
\begin{equation}\label{eq:initial_vector}
t = z_{0}\,, \qquad \lambda(t = z_{0}) = z_{0}\,, \qquad \eta(t = z_0) 
= \dfrac{K_{0}}{1-\theta_{h}}\,,
\end{equation}
with $y(t=z_0;z) = y_v(z)$, and
with the local monodromy exponents being set by 
\begin{align}\label{eq:monoinitvector}
    \vartheta_0 = -1\,\,,\,\,\vartheta_t = \theta_h -1\,\,,\,\,\vartheta_1 = -\sqrt{1+\mu^2}\,\,,\,\,
    \vartheta_\infty = \theta_c + 1\,\,,
\end{align}
where $\theta_{h}$, $\theta_{c}$, $z_{0}$, and $K_{0}$ are given by Eqs.~\eqref{eq:thetas}, \eqref{eq:zee0}, and \eqref{eq:accessory_vector}, respectively. For consistency check, 
the initialization in Eq.~\eqref{eq:initial_vector} with the monodromies in 
Eq.~\eqref{eq:monoinitvector} gives $K(t=z_0) = 0$ as it should from Eq.~\eqref{eq:heun_vector}.
Again, for fixed $\tilde{r}_h$, $\mu$ and $\ell$, the expansion of the PVI $\tau$ function 
is determined up to the local isomonodromies $\vartheta_t,\vartheta_{\infty}$, which depend on the quasinormal frequency, and the composite monodromy $\sigma$.
The pair $(\tilde{\omega}, \sigma)$ can then be computed by solving 
the initial conditions for the PVI $\tau$ function Eq.~\eqref{eq:conditions}, 
which in the case of the vector-type sector, take the following form
\begin{subequations}\label{eq:conditions_vector}
\begin{equation}
    \begin{split}
        \frac{d}{d t}\log \tau(\rho;t)\biggl\vert_{t=z_{0}}  = \dfrac{(1 - \theta_{h})}{2 z_{0}} + \dfrac{(1 - \theta_{h})\sqrt{1+\tilde{\mu}^{2}}}{2(z_{0} - 1)} + K_{0}\,,
    \end{split}
\end{equation}
\begin{equation}
    \begin{split}
        \frac{d}{dt}\biggl[t(t-1)\frac{d}{d t}\log \tau(\rho;t)\biggr]\biggl\vert_{t = z_{0}}  = \frac{1}{2}(\theta_{h} -1)(\theta_{c} - \theta_{h} + 2)\,.
    \end{split}
\end{equation}
\end{subequations}

\section{Numerical results}
\label{sec:5}

\subsection{Results using the isomonodromic method}
\label{sec:5.1}
In this section, we present the numerical computation of the QNM frequencies of the scalar- and vector-type sectors in the small $\tilde{r}_{h}$ black hole regime. First, we implement the initial conditions in Eqs.~\eqref{eq:conditions_scalar} and \eqref{eq:conditions_vector} using the conformal blocks expansion of the PVI $\tau$ function, Eq.~\eqref{eq:nekrasovexpansion}, truncated at $\mathcal{O}(t^{9})$. Then, the resulting transcendental equations are solved in \texttt{Python} by applying a root finding algorithm which employs the Muller's method.

\begin{figure}[!h]
\centering
\includegraphics[width=\linewidth]{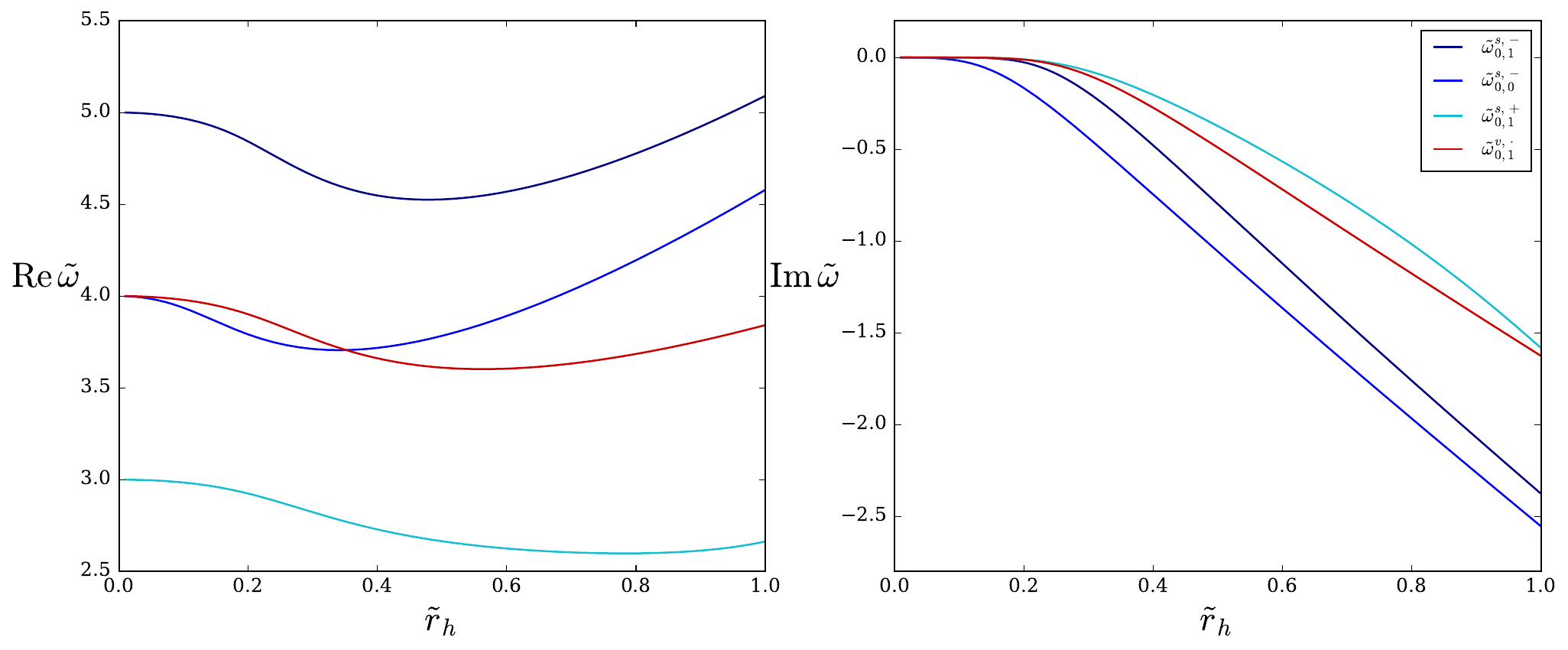}
\caption{Numerical results for the QNMs frequencies as a function of the horizon radius 
for the scalar-type and vector-type modes. The mass of the scalar field is $\mu = 0.001$.}
\label{fig:modes}
\end{figure}

In Fig.~\ref{fig:modes}, we display the QNM frequencies $\tilde{\omega}_{n,\ell}^{i,j}$ as a function of the horizon radius and fixed mass of the field $\tilde{\mu}=0.001$. The index $i$ denotes the type of mode: scalar or vector; $j$ refers to its polarization: electromagnetic polarization $\tilde{\beta}_{+}$ or non-electromagnetic polarization $\tilde{\beta}_{-}$, while $n$ and $\ell$ correspond to the principal quantum number and the angular momentum quantum number, respectively. Within the scalar-type sector, we present the fundamental modes $(n=0)$ for different $\ell$. Namely, the electromagnetic polarization modes $\tilde{\omega}_{0,1}^{s,+}$, the non-electromagnetic polarization modes $\tilde{\omega}_{0,1}^{s,-}$, and the monopole modes $\tilde{\omega}_{0,0}^{s,-}$. We recall the reader that the initial conditions for the electromagnetic and non-electromagnetic polarization, as well as the monopole modes, are the same. For vector-type modes, we compute the fundamental modes for $\ell=1$, $\tilde{\omega}_{0,1}^{v,\cdot}$. Finally, in the limit $\tilde{r}_{h} \to 0$, our numerical results coincide with the analytic formula for the normal modes frequencies found in Eq.~$(61)$ of \cite{Lopes:2024ofy}.

\subsection{Comparison between isomonodromic method and 
numerical integration}
\label{sec:5.2}

Here, we perform the comparison between the isomonodromic 
method results and the numerical integration method. This 
last method is based on performing the numerical integration 
of the radial differential equation, see 
Refs.~\cite{Pani:2013pma, Chandrasekhar:1975zza, 
Konoplya:2008rq, Zhidenko:2009zx, 
Rosa:2011my}. The objective of this 
comparison is to see the relative accuracy of both approaches.
The isomonodromic method is known to be accurate for small 
$\tilde{r}_h$, while the numerical integration method is 
known to be 
accurate for large $\tilde{r}_h$. 
Therefore, this comparison is able 
to show at what values of $\tilde{r}_h$, one of the 
approaches seems to deviate from the expected value.

\begin{figure}[!htb]
\centering
\begin{subfigure}[h]{0.49\linewidth}
     \centering
    \includegraphics[width=0.95\linewidth]{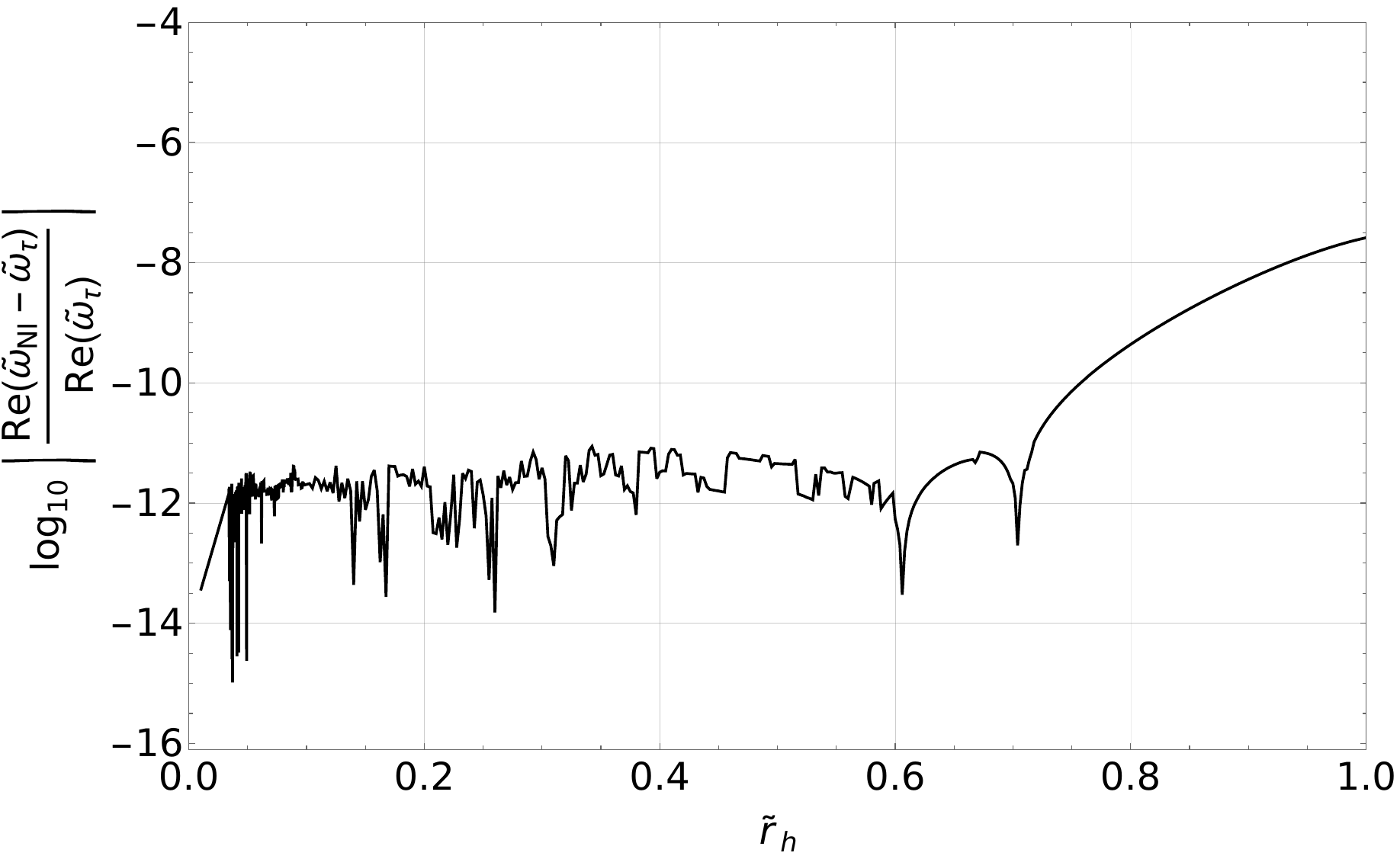}
    \caption{Real part}
\end{subfigure}%
\begin{subfigure}[h]{0.49\linewidth}
     \centering
    \includegraphics[width=0.95\linewidth]{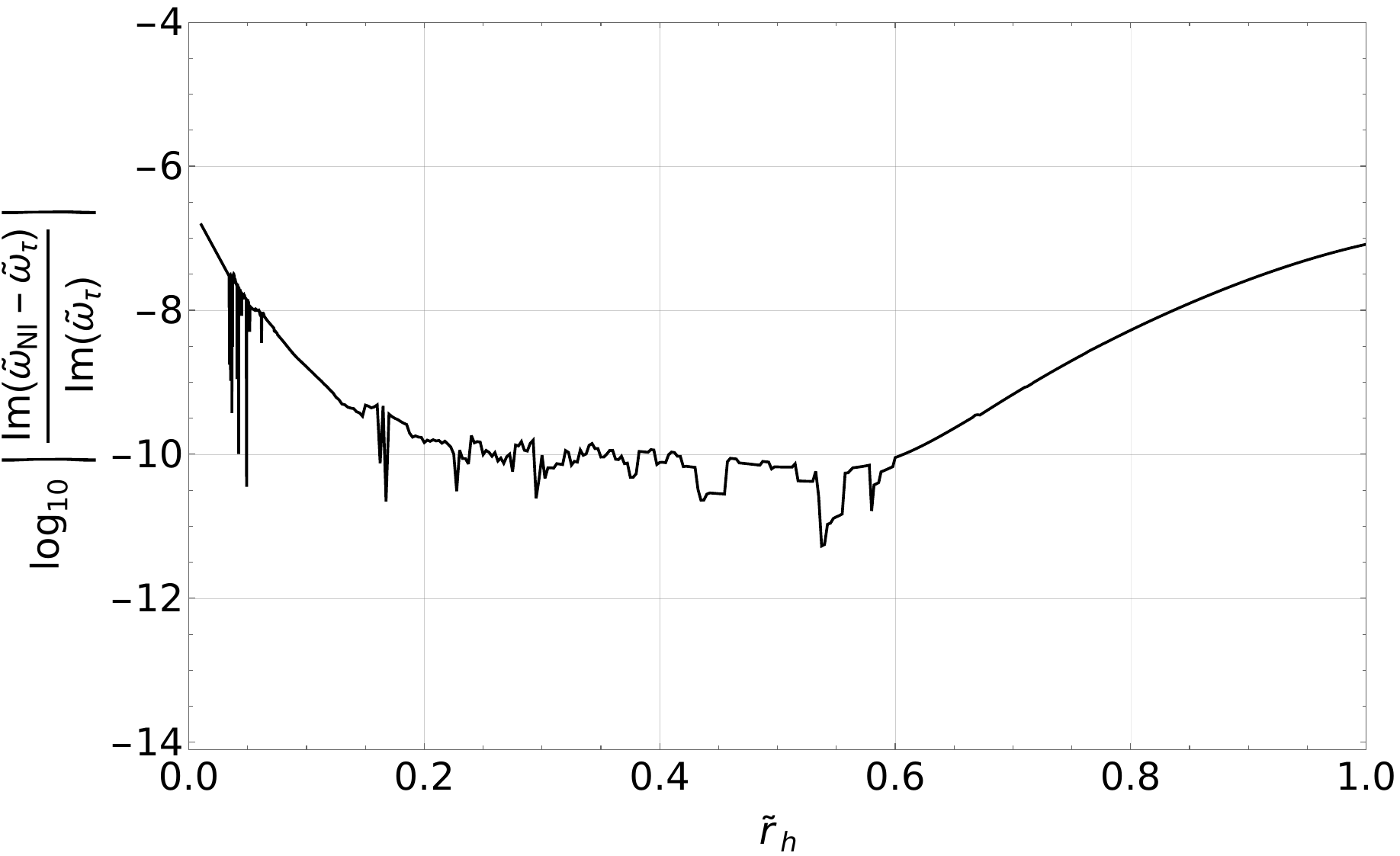}
    \caption{Imaginary part}
\end{subfigure}
    \caption{Relative difference between the real and imaginary parts of the fundamental quasinormal mode frequencies 
    provided by the isomonodromic method 
    $\tilde{\omega}_\tau$ and provided by the 
    numerical integration method $\tilde{\omega}_{\mathrm{NI}}$, 
    in logarithmic scale, for the scalar-type electromagnetic polarization $\beta_+$.}
    \label{fig:diffbetaplus}
\end{figure}

For the numerical integration method, the radial differential 
equation is integrated starting close to the horizon at 
$\tilde{r}_i = 1.01\tilde{r}_h$ up to a radius $\tilde{r}=\tilde{r}_m$. 
For the initial conditions, one 
expands in series the solution near the horizon, assuming the 
boundary conditions at $\tilde{r}_h$, and one considers the 
coefficients up to sixth order given by the 
recurrence relations of the differential equation. 
Then, one uses as 
initial conditions the value at $\tilde{r}_i$ 
of the solution and its first derivative provided by this 
expansion. Similarly, one performs another integration, this time starting close to infinity at $\tilde{r}_f = 10^{11}\tilde{r}_h$ down to $\tilde{r}=\tilde{r}_m$. The initial conditions for this integration are obtained by expanding the solution in series up to sixth order in $1/\tilde{r}$, assuming the boundary conditions at infinity, with coefficients determined by the recurrence relations of the differential equation. One then matches both solutions at $\tilde{r}=\tilde{r}_m$, which amounts to finding the quasinormal mode frequency that makes their Wronskian vanish at $\tilde{r}=\tilde{r}_m$. Since the Wronskian is independent of the radius, $\tilde{r}=\tilde{r}_m$ can be chosen without loss of generality. Throughout the numerics, we have used $\tilde{r}_m = 0.67\tilde{r}_h$. 
Furthermore, in order to compute the frequencies, we initialized the root finder to the 
frequencies obtained by the isomonodromy method.

\begin{figure}[!htb]
\begin{subfigure}[h]{0.49\linewidth}
     \centering
    \includegraphics[width=0.95\linewidth]{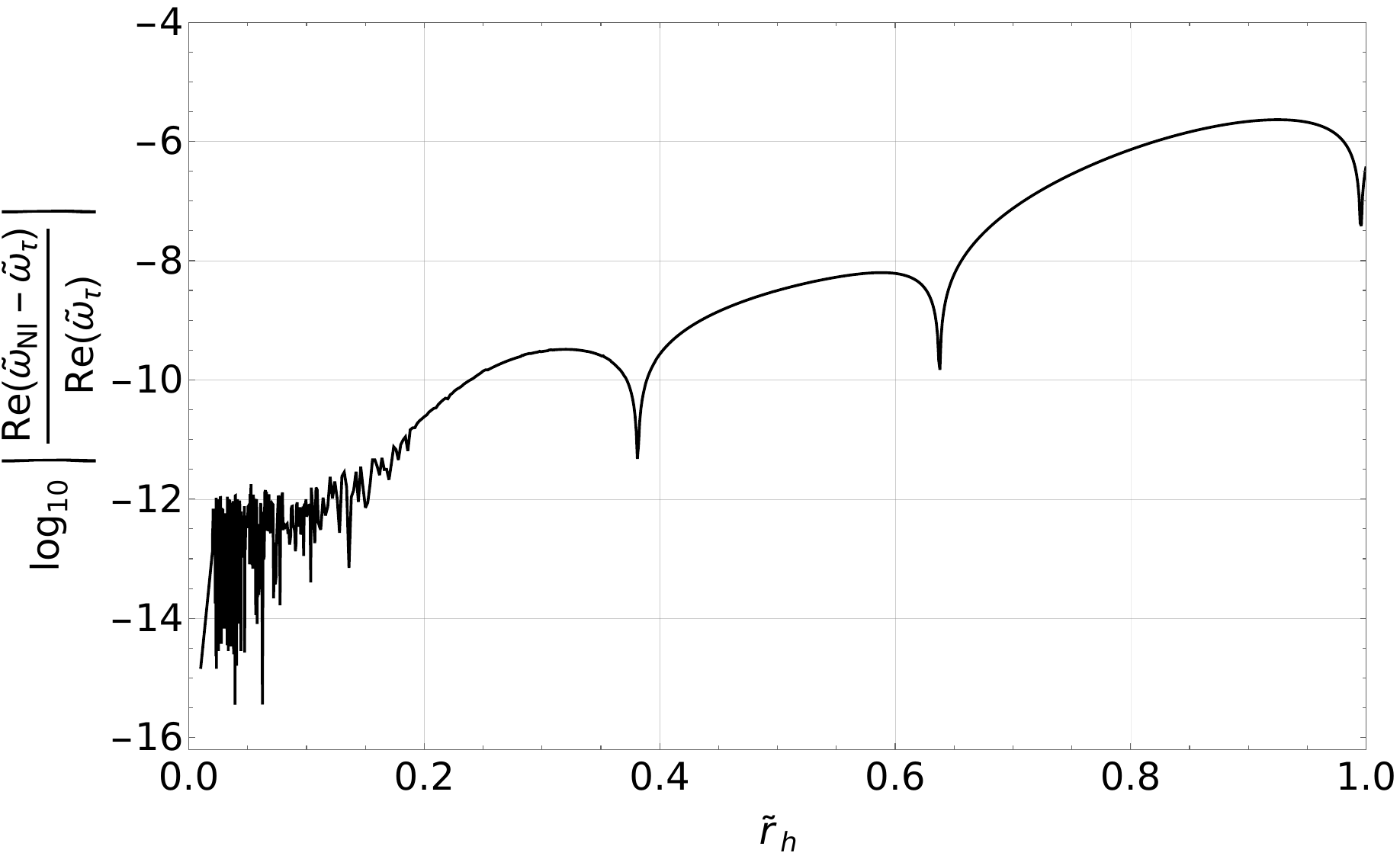}
    \caption{Real part}
\end{subfigure}%
\begin{subfigure}[h]{0.49\linewidth}
     \centering
    \includegraphics[width=0.95\linewidth]{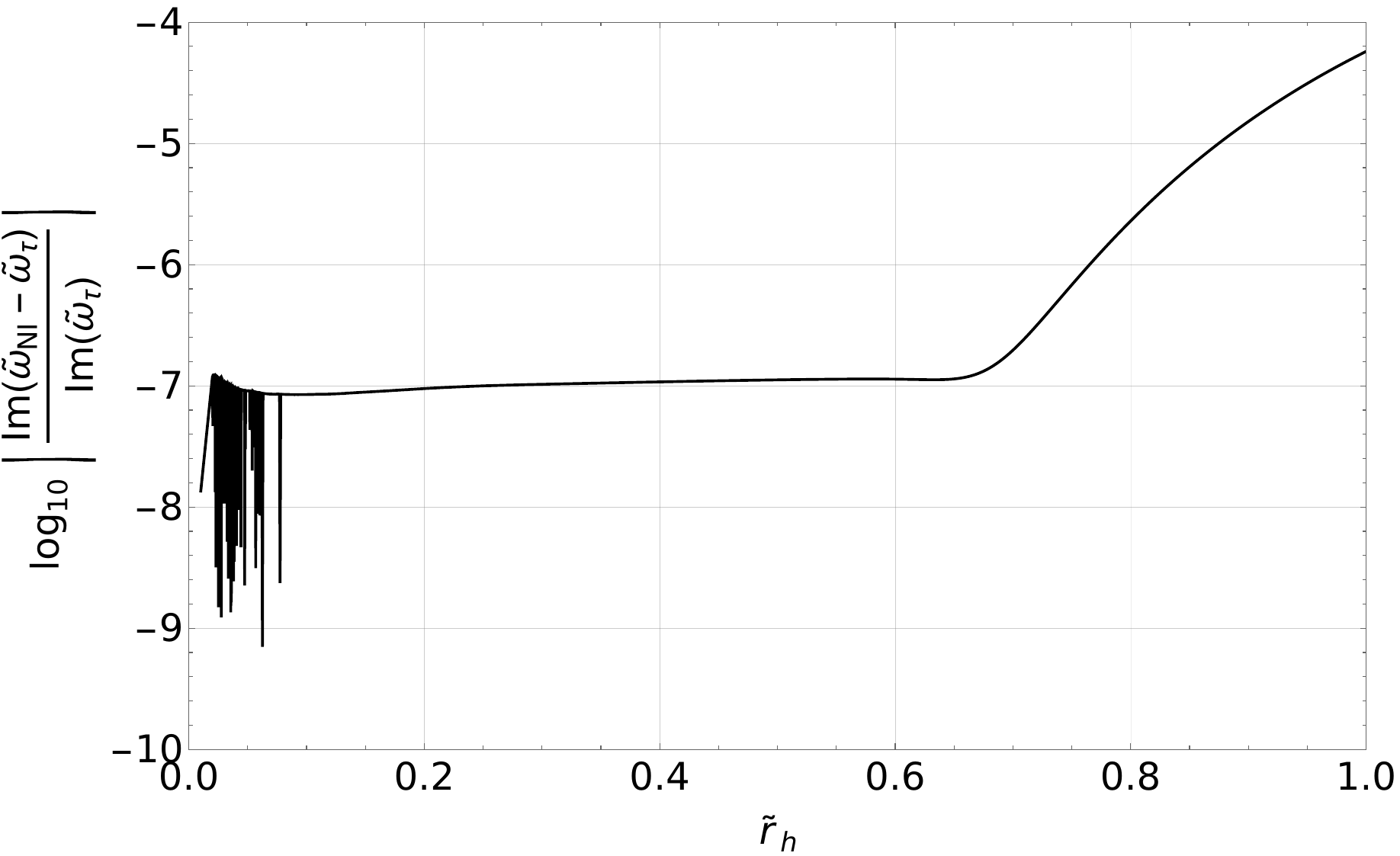}
    \caption{Imaginary part}
\end{subfigure}
    \caption{Relative difference between the real and imaginary parts of the 
    fundamental quasinormal mode frequencies 
    provided by the isomonodromic method 
    $\tilde{\omega}_\tau$ and provided by the 
    numerical integration method $\tilde{\omega}_{\mathrm{NI}}$, 
    in logarithmic scale, for the scalar-type non-electromagnetic polarization $\beta_-$.}
    \label{fig:diffbetaminus}
\end{figure}

\begin{figure}[!htb]
\centering
\begin{subfigure}[h]{0.49\linewidth}
     \centering
    \includegraphics[width=0.95\linewidth]{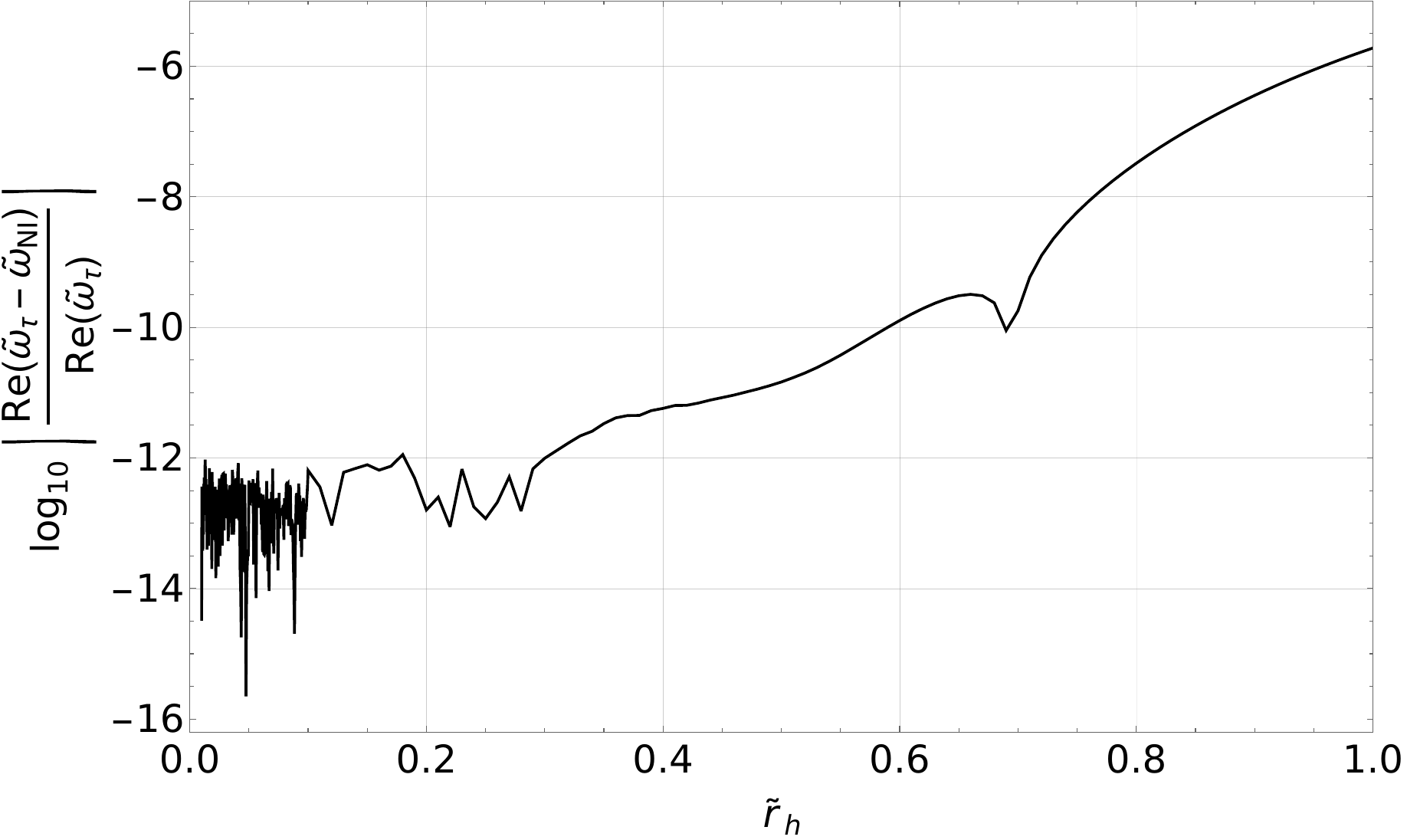}
    \caption{Real part}
\end{subfigure}%
\begin{subfigure}[h]{0.49\linewidth}
     \centering
    \includegraphics[width=0.95\linewidth]{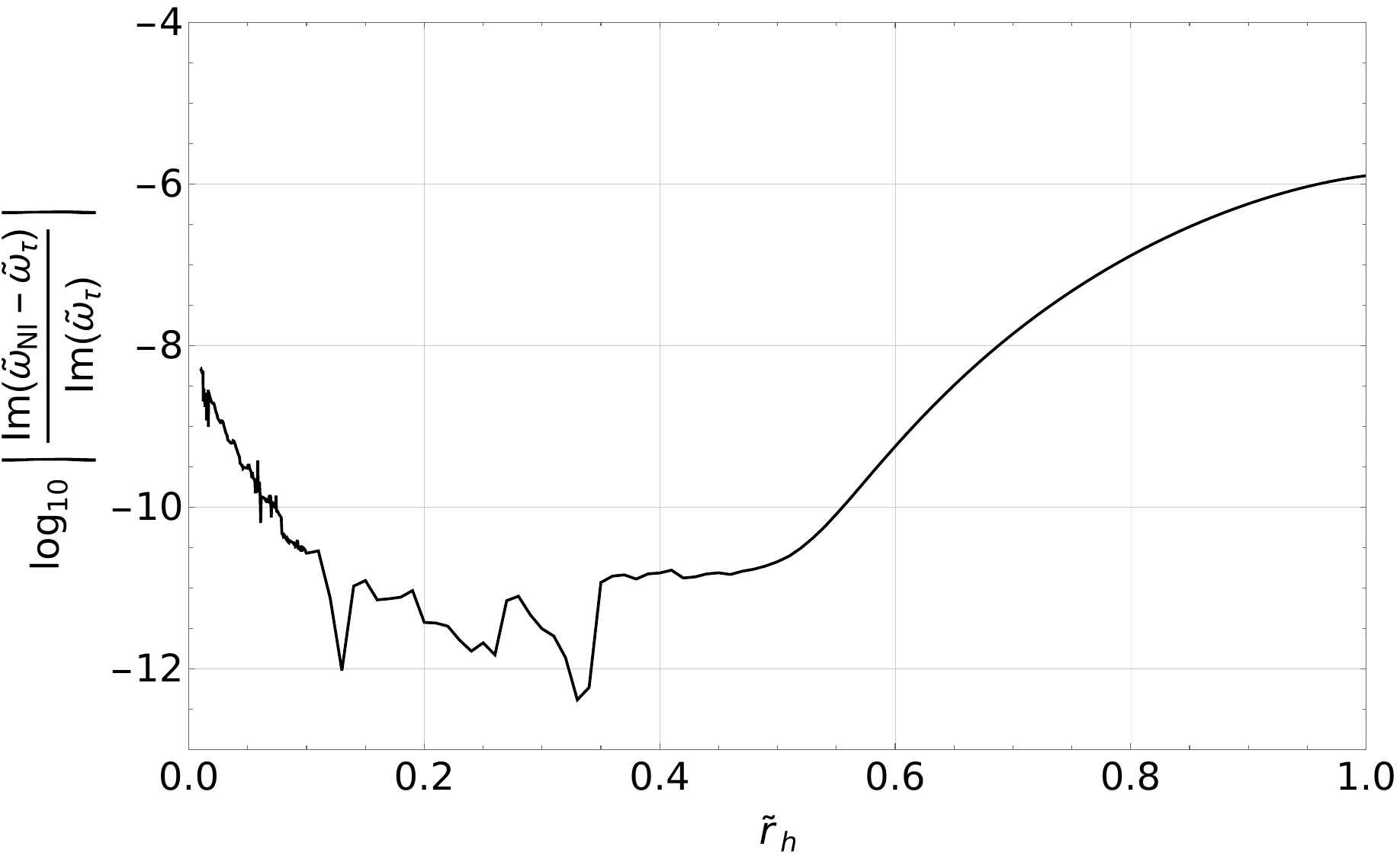}
    \caption{Imaginary part}
\end{subfigure}
    \caption{Relative difference between the real and imaginary parts of the fundamental quasinormal mode frequencies 
    provided by the isomonodromic method 
    $\tilde{\omega}_\tau$ and provided by the 
    numerical integration method $\tilde{\omega}_{\mathrm{NI}}$, 
    in logarithmic scale, for the monopole mode.}
    \label{fig:diffmonopole}
\end{figure}

\begin{figure}[!htb]
\centering
\begin{subfigure}[h]{0.49\linewidth}
     \centering
    \includegraphics[width=0.95\linewidth]{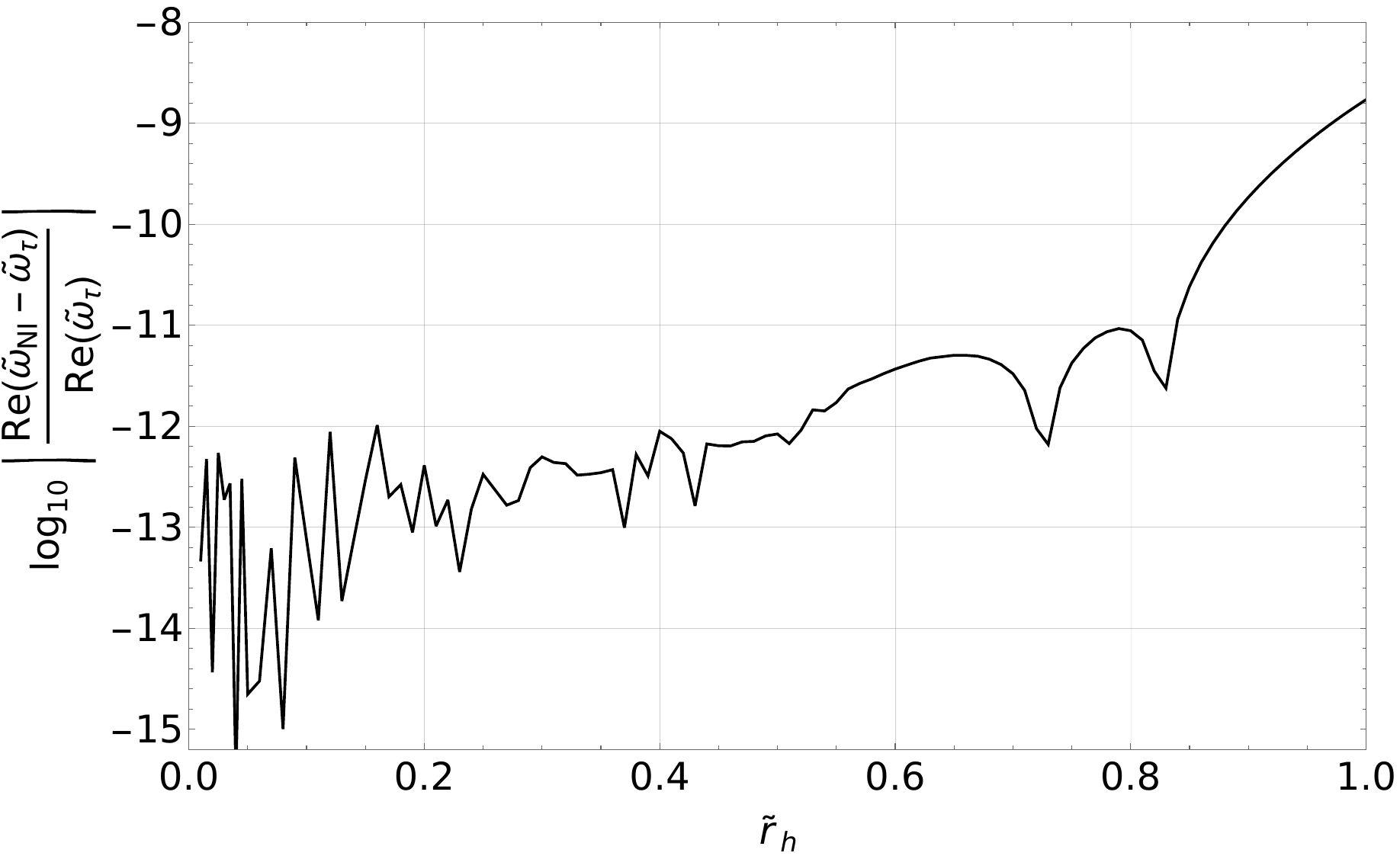}
    \caption{Real part}
\end{subfigure}%
\begin{subfigure}[h]{0.49\linewidth}
     \centering
    \includegraphics[width=0.95\linewidth]{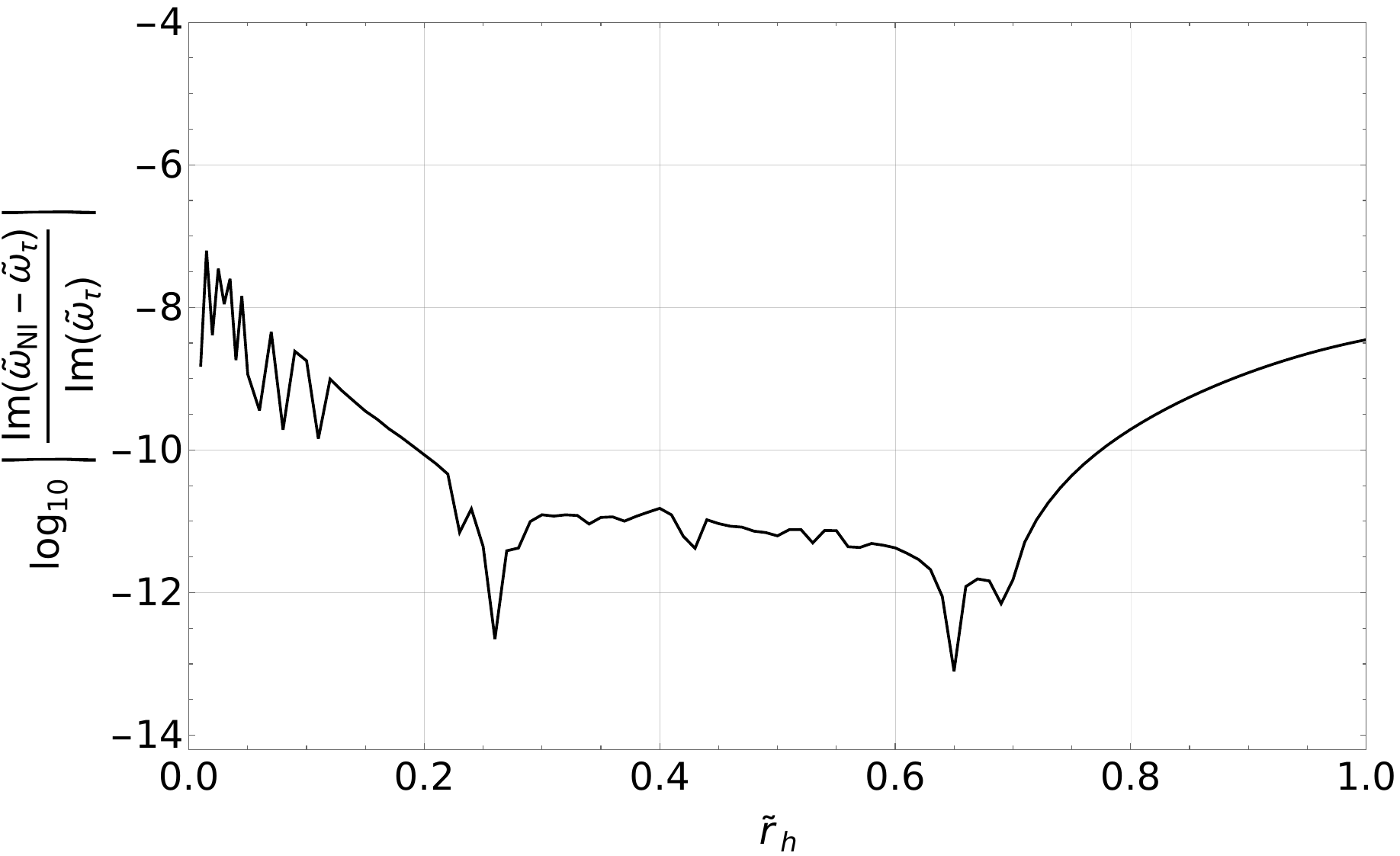}
    \caption{Imaginary part}
\end{subfigure}
    \caption{Relative difference between the real and imaginary parts of the fundamental quasinormal mode frequencies 
    provided by the isomonodromic method 
    $\tilde{\omega}_\tau$ and provided by the 
    numerical integration method $\tilde{\omega}_{\mathrm{NI}}$, 
    in logarithmic scale, for the vector-type mode.}
    \label{fig:diffvector}
\end{figure}

The relative difference between the quasinormal mode 
frequencies of both methods are found in Fig.~\ref{fig:diffbetaplus} for 
the scalar-type electromagnetic polarization mode $\beta_+$,
in Fig.~\ref{fig:diffbetaminus} for the scalar-type non-electromagnetic 
polarization mode $\beta_-$, in Fig.~\ref{fig:diffmonopole} for the monopole 
mode, and in Fig.~\ref{fig:diffvector} for the vector-type mode. 
Overall, the two methods agree very well, having very small relative differences 
of up to $10^{-2}\%$, which is the maximum relative difference for the 
imaginary part of the scalar-type non-electromagnetic polarization. However, 
the behaviour of the relative differences seems to show the regimes of 
accuracy of both methods. 

A common trend in all the relative differences is that the 
real part given by both methods agrees very well for very 
small $r_h$, having relative differences around the order of 
$10^{-13}$ for all polarizations. However, the relative 
differences start to increase for increasing $r_h$, 
up to $10^{-8}$ for the scalar-type electromagnetic
polarization $\beta_+$, up to $10^{-6}$ for the scalar-type 
non-electromagnetic polarization $\beta_-$, up to 
$10^{-6}$ for the monopole mode and up to $10^{-9}$ for the 
vector-type mode, when the horizon radius is close to unity.
Since it is expected for the numerical integration method 
to do well for the horizon radius close to unity, then 
our results show that the isomonodromic method starts to 
deviate from the expected frequency value, for horizon radius 
higher than unity.

Regarding the imaginary part, there is an interesting 
behaviour. The relative differences start from a 
value around $10^{-7}$ for the scalar-type electromagnetic 
polarization, $10^{-7}$ for the scalar-type non-electromagnetic 
polarization, $10^{-8}$ for the monopole and 
$10^{-8}$ for the vector-type mode. Then, the relative 
differences decrease until they reach an intermediate value 
of $\tilde{r}_h$ and increase afterwards reaching a value 
at $\tilde{r}_h$ close to unity of $10^{-7}$ for the scalar-type 
electromagnetic polarization, $10^{-4}$ for the 
scalar-type non-electromagnetic polarization, $10^{-6}$ 
for the monopole mode and $10^{-8}$ for the vector-type 
polarization. The fact that the relative differences are 
higher for very low $r_h$, assuming that the isomonodromic method 
is accurate for this range, means that the numerical 
integration method is not accurate to compute the imaginary 
part of the frequency for small $r_h$. But the relative 
differences are also higher for $r_h$ close to unity, where 
it accompanies the trend of the real part of the frequency, 
as the isomonodromic method starts to have less accuracy.

Therefore, the comparison of the quasinormal mode frequencies
agree with the fact that the isomonodromic method is 
accurate for small values of the horizon radius, while the 
numerical integration method is accurate for higher values of 
the horizon radius, and also accurate to capture the real part of the 
frequency for small values of 
the horizon radius. However, a point can be made that 
maybe the overall accuracy of both methods may be low in 
capturing the imaginary part for very small horizon radius. 
Indeed, in this case, the imaginary part is very close to 
zero and so it may be plagued with numerical error.

\clearpage

\section{Conclusions}
\label{sec:6}
In this paper, the quasinormal modes of the Proca field in a Schwarzschild-AdS$_{5}$ spacetime were obtained numerically, using the isomonodromy method.
The Proca field was decomposed into scalar-type and vector-type components by using the decomposition 
in scalar and vector spherical harmonics. While the components of the scalar-type are coupled, the vector-type component is completely decoupled. In turn, we introduce the FKKS ansatz, which in the Schwarzschild limit gives a transformation 
which separates the scalar-type modes. 

In the scalar-type sector, the radial differential equation contains five regular singular points 
, with one singularity arising due to the separation parameter $\beta$. Since the Frobenius solutions around this singularity have characteristic exponents $(0,2)$, we have assumed that it is an apparent singularity, which implies that the radial ODE can be interpreted as a deformed Heun equation. Thus, the initial conditions for the isomonodromic flow 
are defined at the point where the deformed Heun equation 
coincides with the radial differential equation of the scalar-type sector. 

In contrast, the radial differential equation of the vector-type sector has only four regular singular points. 
Hence, the extra singularity introduced by the isomonodromic deformation of the Fuchsian system serves an auxiliary role in solving the problem. Once the initial conditions are imposed
, the apparent singularity merges with one of the other singular points, thus reducing the deformed Heun equation to a Heun equation, which can be mapped to the radial ODE of the vector-type sector. 

Interestingly, the initial conditions of the scalar-type and vector-type  are given by the PVI $\tau$ function, indicating that their dynamical systems evolve with the same Hamiltonian, and therefore radial ODEs correspond to different points in the phase space of the isomonodromic flow $(\lambda(t),\eta(t))$.

The quasinormal modes were then obtained using the isomonodromy method for small 
event horizon radius. The results were then compared with the numerical integration method. 
It is found that both methods have an overall very good agreement, up to $10^{-2}\%$. However, the behaviour of the relative differences corroborates the expectation that the isomonodromy method 
has better accuracy than the numerical integration method for very small horizon radius, while 
numerical integration method starts to have better accuracy for intermediate horizon radius. 
Since one does not know exactly the value of the quasinormal modes, one cannot state with 
full certainty that the expectation is indeed correct. One would need to find an analytical 
expression for small horizon radius in order to compare with these numerical results.

\acknowledgments
J.B.A. acknowledges financial support from the Funda\c{c}\~{a}o para a Ci\^{e}ncia e a Tecnologia (FCT) - Portugal through the research project No. 10.54499/2022.03702.PTDC (GENIDE). 
T.V.F. acknowledges financial support from the Funda\c{c}\~{a}o para a Ci\^{e}ncia e a 
Tecnologia (FCT) through the project No. UIDB/00099/2025, namely through the grant FCT no. RD1415.

\appendix
\section{Painlev\'{e} VI $\tau$ function}
\label{appendix:A}

The series expansion of the PVI $\tau$ function, written in \cite{Gamayun:2012ma,Gamayun:2013auu}, near $t=0$ is given by
\begin{equation}\label{eq:nekrasovexpansion}
    \tau(t)=\sum_{n\in\mathbb{Z}}C(\vec{\vartheta},\sigma+2n)s^nt^{\tfrac{1}{4}((\sigma+2n)^2-\vartheta_{0}^{2}-\vartheta_{t}^{2})}\mathcal{B}(\vec{\vartheta},\sigma+2n;t)\,,
\end{equation}
where $\vec{\vartheta}=\lbrace\vartheta_{0},\vartheta_{t},\vartheta_{1},\vartheta_{\infty}\rbrace$ are the local monodromy exponents, and the parameters $\sigma$, $s$ are two integration constants. The structure constants $C(\vec{\vartheta},\sigma)$ are expressed in terms of Barnes' functions
\begin{equation}
    C(\vec{\vartheta},\sigma)=\frac{\prod_{\alpha,\beta=\pm}
	   G(1+\tfrac{1}{2}( \vartheta_{1}+\alpha\vartheta_{\infty}+\beta\sigma))
	   G(1+\tfrac{1}{2}( \vartheta_{t}+\alpha\vartheta_{0}+\beta\sigma))}{G(1+\sigma)G(1-\sigma)},
\end{equation}
and $\mathcal{B}(\vec{\vartheta},\sigma;t)$ is a power series in $t$ which coincides with the $c=1$ Virasoro conformal blocks, and is explicitly given by
\begin{subequations}
\begin{equation}\label{eq:CB}
    \mathcal{B}(\vec{\vartheta},\sigma+2n;t) = (1-t)^{\tfrac{1}{2}\vartheta_{t}\vartheta_{1}}\sum_{\lambda,\mu\in \mathbb{Y}}{\cal B}_{\lambda,\mu}(\vec{\vartheta},\sigma+2n) t^{\vert\lambda\vert + \vert\mu\vert},
\end{equation}
\begin{equation}
\begin{split}
  {\cal B}_{\lambda,\mu}(\vec{\vartheta},\sigma) &=
  \prod_{(i,j)\in\lambda}\frac{((\vartheta_t+\sigma+2(i-j))^2-\vartheta_0^2)
    ((\vartheta_1+\sigma+2(i-j))^2-\vartheta_\infty^2)}{16h_\lambda^2(i,j)
    (\lambda'_j-i+\mu_i-j+1+\sigma)^2} \\
  &\times \prod_{(i,j)\in\mu}\frac{((\vartheta_t-\sigma+2(i-j))^2-\vartheta_0^2)
    ((\vartheta_1-\sigma+2(i-j))^2-\vartheta_\infty^2)}{16h_\lambda^2(i,j)
    (\mu'_j-i+\lambda_i-j+1-\sigma)^2}\,,
\end{split}
\end{equation}
\end{subequations}
where the sum is over pairs $(\lambda,\mu)$ of Young diagrams on $\mathbb{Y}$. The size of the diagram is given by the number of boxes in it, thus $\vert\lambda\vert$ (or analogously, $\vert\mu\vert$). Furthermore, for each box situated at $(i,j)$ in $\lambda$, $\lambda_i$ is the number of boxes at row $i$ of $\lambda$ and $\lambda'_j$ is the number of boxes at column $j$ of $\lambda$; $h(i,j)=\lambda_i+\lambda'_j-i-j+1$ is the hook length of the box at $(i,j)$.

The parameter $s$ can be determined in terms of the monodromy matrices $\lbrace \sigma, \sigma_{t1} \rbrace$ from the formula $(3.48a)$ in \cite{ILP2016}:
\begin{equation}\label{eq:ess}
    \begin{split}
        \sin^2\pi\sigma\cos\pi\sigma_{t1} &= \cos\pi\vartheta_{0}\cos\pi\vartheta_{\infty} + \cos\pi\vartheta_{t}\cos\pi\vartheta_{1}\\
        &- \cos\pi\sigma(\cos\pi\vartheta_{0}\cos\pi\vartheta_{1} + \cos\pi\vartheta_{t}\cos\pi\vartheta_{\infty}) \\
        &- \frac{1}{2}(\cos\pi\vartheta_{\infty} - \cos\pi(\vartheta_{1} - \sigma))
        (\cos\pi\vartheta_{0} - \cos\pi(\vartheta_{t} - \sigma))s \\  
        &- \frac{1}{2}(\cos\pi\vartheta_{\infty} - \cos\pi(\vartheta_{1} + \sigma))
        (\cos\pi\vartheta_{0} - \cos\pi(\vartheta_{t} + \sigma))s^{-1}\,.  
    \end{split}
\end{equation}
For $t$ sufficiently close to zero\footnote{Analogous expansions of $\tau(t)$ around other critical points $t = \lbrace 1,\infty \rbrace$ can be obtained by applying a suitable transformation of parameters, see for instance \cite{Jimbo:1982}.}, and generic monodromy parameters in the sense that
\begin{equation}\label{eq:monoconds}
    \sigma \notin \mathbb{Z}, \qquad \sigma \pm \vartheta_{0} \pm \vartheta_{t} \notin \mathbb{Z}, \qquad \sigma \pm \vartheta_{1} \pm \vartheta_{\infty} \notin \mathbb{Z},
\end{equation}
we have
\begin{equation}\label{eq:asymp_tau}
    \begin{split}
        \tau(t) = C_{0}t^{\tfrac{1}{4}(\sigma^{2}-\vartheta^{2}_{0}-\vartheta^{2}_{t})}(1-t)^{\tfrac{1}{2}\vartheta_{t}\vartheta_{1}}\biggl\{1 &+\bigg[\frac{\vartheta_{t}\vartheta_{1}}{2}+\frac{(\vartheta^{2}_{0}-\vartheta^{2}_{t}-\sigma^{2})(\vartheta^{2}_{\infty}-\vartheta^{2}_{1}-\sigma^{2})}{8\sigma^{2}} \\
        &-\frac{(\vartheta^{2}_{0}-(\vartheta_{t}-\sigma)^{2})(\vartheta^{2}_{\infty}-(\vartheta_{1}-\sigma)^{2})}{16\sigma^{2}(1+\sigma)^{2}}\kappa\,t^{\sigma}\\
        &- \frac{(\vartheta^{2}_{0}-(\vartheta_{t}+\sigma)^{2})(\vartheta^{2}_{\infty}-(\vartheta_{1}+\sigma)^{2})}{16\sigma^{2}(1-\sigma)^{2}}\frac{1}{\kappa\,t^{\sigma}}\bigg]t + \cdots\biggr\},
    \end{split}
\end{equation}
where $0 < \mathrm{Re}\,\sigma < 1$, $C_{0}$ is a constant independent of $t$, and $\kappa$ is a known function of the monodromy parameters:
\begin{equation}\label{eq:kappa}
    \begin{split}
        \kappa=s\frac{\Gamma^2(1-\sigma)}{\Gamma^2(1+\sigma)}
        &\frac{\Gamma(1+\tfrac{1}{2}(\vartheta_{t} + \vartheta_{0} + \sigma))
        \Gamma(1+\tfrac{1}{2}(\vartheta_{t} - \vartheta_{0} + \sigma))}{
        \Gamma(1+\tfrac{1}{2}(\vartheta_{t} + \vartheta_{0} - \sigma))
        \Gamma(1+\tfrac{1}{2}(\vartheta_{t} - \vartheta_{0} - \sigma))}\\
        &\qquad \frac{\Gamma(1+\tfrac{1}{2}(\vartheta_{1} + \vartheta_{\infty} +\sigma))
        \Gamma(1+\tfrac{1}{2}(\vartheta_{1} - \vartheta_{\infty} + \sigma))}{
        \Gamma(1+\tfrac{1}{2}(\vartheta_{1} + \vartheta_{\infty} - \sigma))
        \Gamma(1+\tfrac{1}{2}(\vartheta_{1} - \vartheta_{\infty} - \sigma))}.
\end{split}
\end{equation}

\section{Comparison between the decoupled scalar-type modes and the 
numerical integration of the coupled system}
\label{appendix:B}

The two scalar-type degrees of freedom of the Proca field are described by the coupled system of Eqs.~\eqref{eq:u1fourier} and~\eqref{eq:u2fourier}. We have shown that the FKKS ansatz decouples this system, leading to Eq.~\eqref{eq:equationR} with $\beta$ given by  Eq.~\eqref{eq:betapm}. Here, we compare the results for the quasinormal mode frequencies obtained from numerically integrating the decoupled system with those obtained from integrating the coupled one.

\begin{figure}[!htb]
\centering
\begin{subfigure}[h]{0.49\linewidth}
     \centering
    \includegraphics[width=0.95\linewidth]{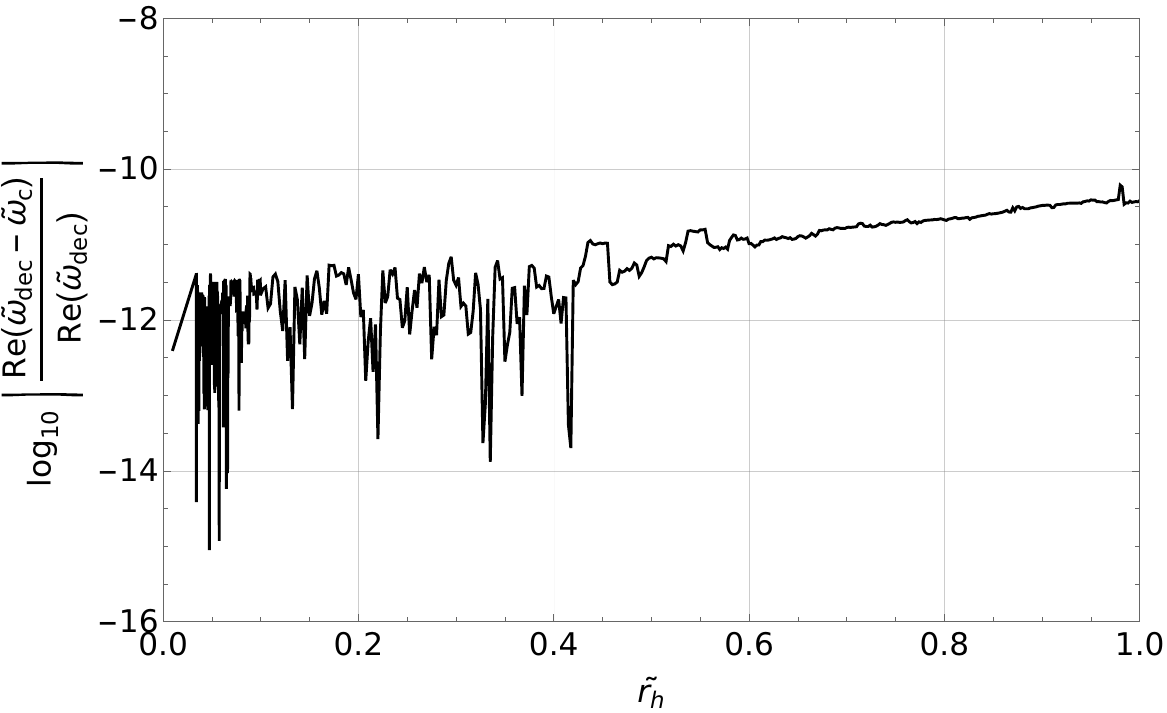}
    \caption{Real part}
\end{subfigure}%
\begin{subfigure}[h]{0.49\linewidth}
     \centering
    \includegraphics[width=0.95\linewidth]{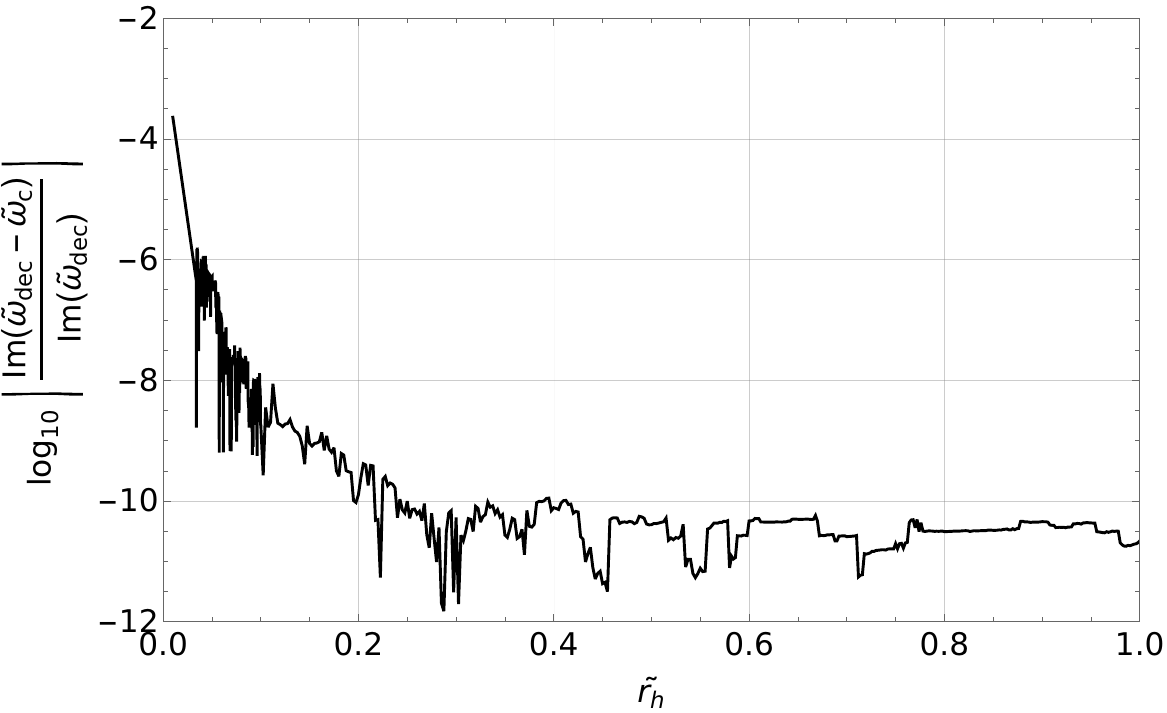}
    \caption{Imaginary part}
\end{subfigure}
    \caption{Relative difference between the real and imaginary parts of the fundamental quasinormal mode frequencies obtained from integrating the coupled system, $\tilde{\omega}_{\mathrm{c}}$, and from integrating the decoupled system, $\tilde{\omega}_{\mathrm{dec}}$, in logarithmic scale, for the scalar-type electromagnetic polarization $\beta_+$.}
    \label{fig:diffdecoupledplus}
\end{figure}

The numerical integration of the coupled system proceeds in a similar way to that described in Section~\ref{sec:5.2}. The initial values for the integrations are obtained from expanding the coupled system near the horizon and near infinity, according to the boundary conditions. The coefficients of the expansions are determined recursively by equating each expansion order. This time, for each integration, there are two free coefficients multiplying the leading-order behaviour of $u_1$ and $u_2$. One needs to choose a suitable orthonormal basis for these coefficients, and perform an integration for each of the two elements of the basis. Thus, in total, one performs four integrations --- two starting from $\tilde{r}=\tilde{r}_i$ up to $\tilde{r}=\tilde{r}_m$, for each basis element, and the other two starting from $\tilde{r}=\tilde{r}_f$ down to $\tilde{r}=\tilde{r}_m$, for each basis element. The quasinormal mode frequencies are then obtained by minimizing the Wronskian of these four solutions at $\tilde{r}=\tilde{r}_m$. All of the numerical values used here were the same as those used in the main text.

\begin{figure}[!htb]
\centering
\begin{subfigure}[h]{0.49\linewidth}
     \centering
    \includegraphics[width=0.95\linewidth]{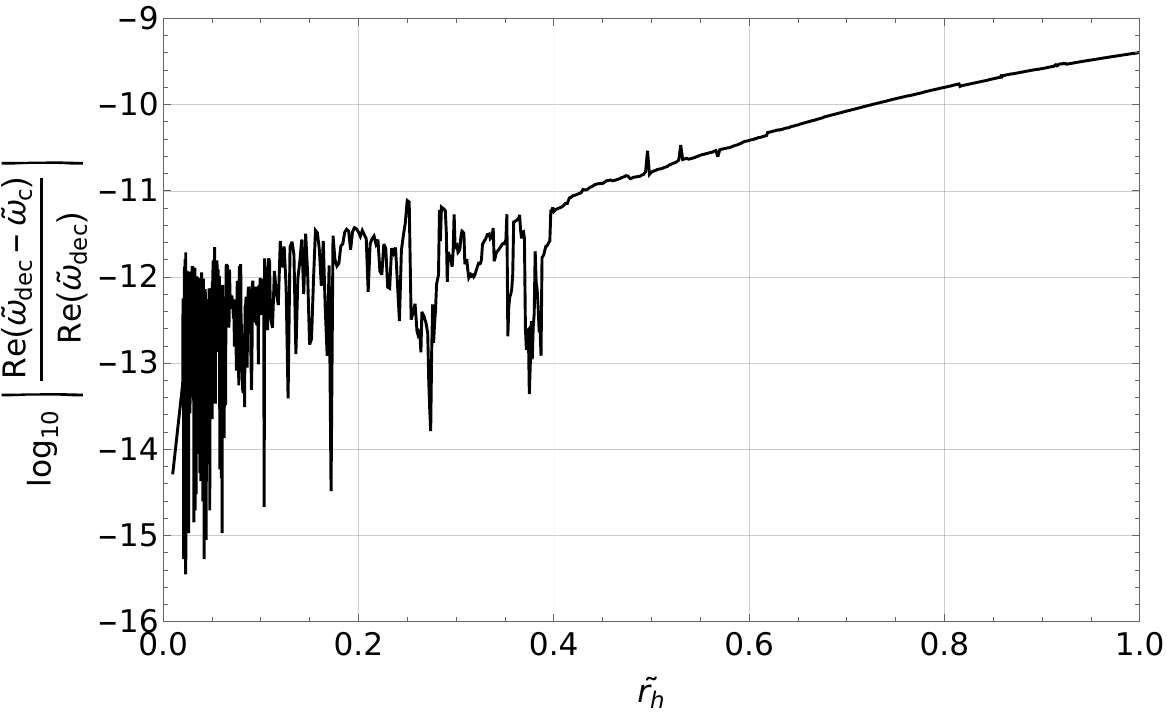}
    \caption{Real part}
\end{subfigure}%
\begin{subfigure}[h]{0.49\linewidth}
     \centering
    \includegraphics[width=0.95\linewidth]{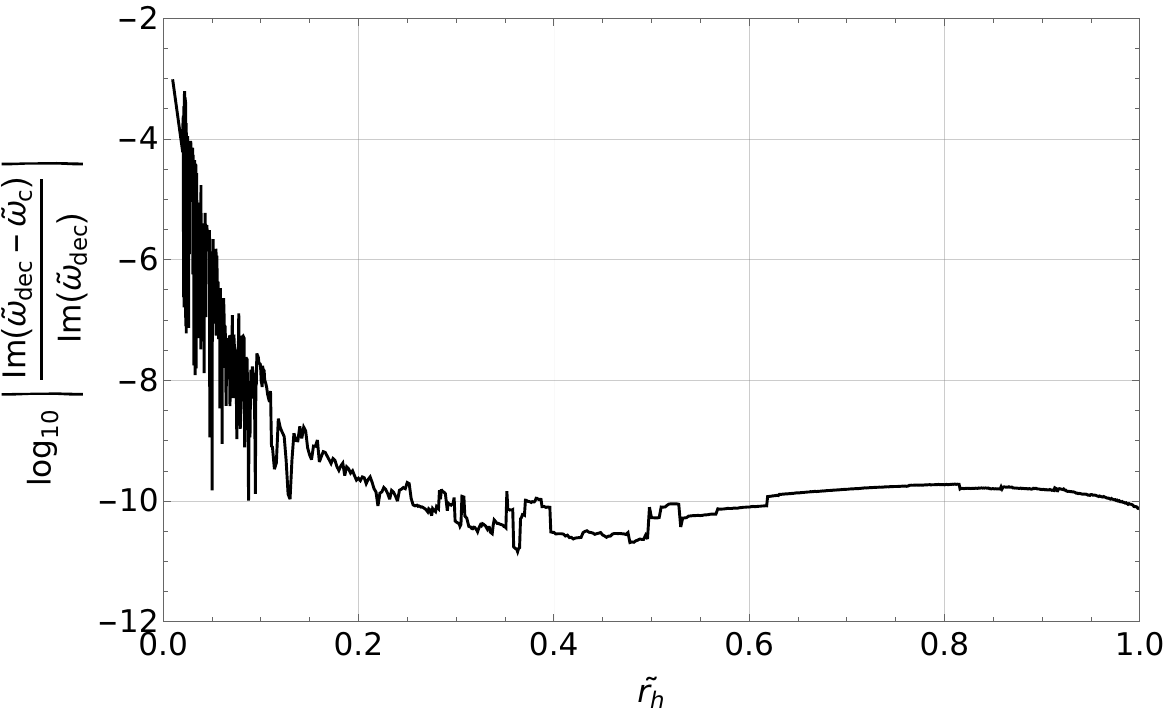}
    \caption{Imaginary part}
\end{subfigure}
    \caption{Relative difference between the real and imaginary parts of the fundamental quasinormal mode frequencies obtained from integrating the coupled system, $\tilde{\omega}_{\mathrm{c}}$, and from integrating the decoupled system, $\tilde{\omega}_{\mathrm{dec}}$, in logarithmic scale, for the scalar-type non-electromagnetic polarization $\beta_-$.}
    \label{fig:diffdecoupledminus}
\end{figure}

In Figs.~\ref{fig:diffdecoupledplus} and~\ref{fig:diffdecoupledminus} we show the relative difference between the quasinormal mode frequencies obtained from integrating the coupled and decoupled systems. The deviations do not seem to depend strongly on the polarization. The real part of the frequencies agrees between the two methods, with a relative deviation of around $10^{-9}$ at most. For black holes with $\tilde{r}_h \gtrsim 0.1$, the imaginary parts of the frequencies show only minor differences between the two methods. For smaller black holes, however, these deviations can reach up to $10^{-3}$, hinting for low reliability of the numerical integration method in this regime.

\begin{figure}[!htb]
\centering
\begin{subfigure}[h]{0.49\linewidth}
     \centering
    \includegraphics[width=0.95\linewidth]{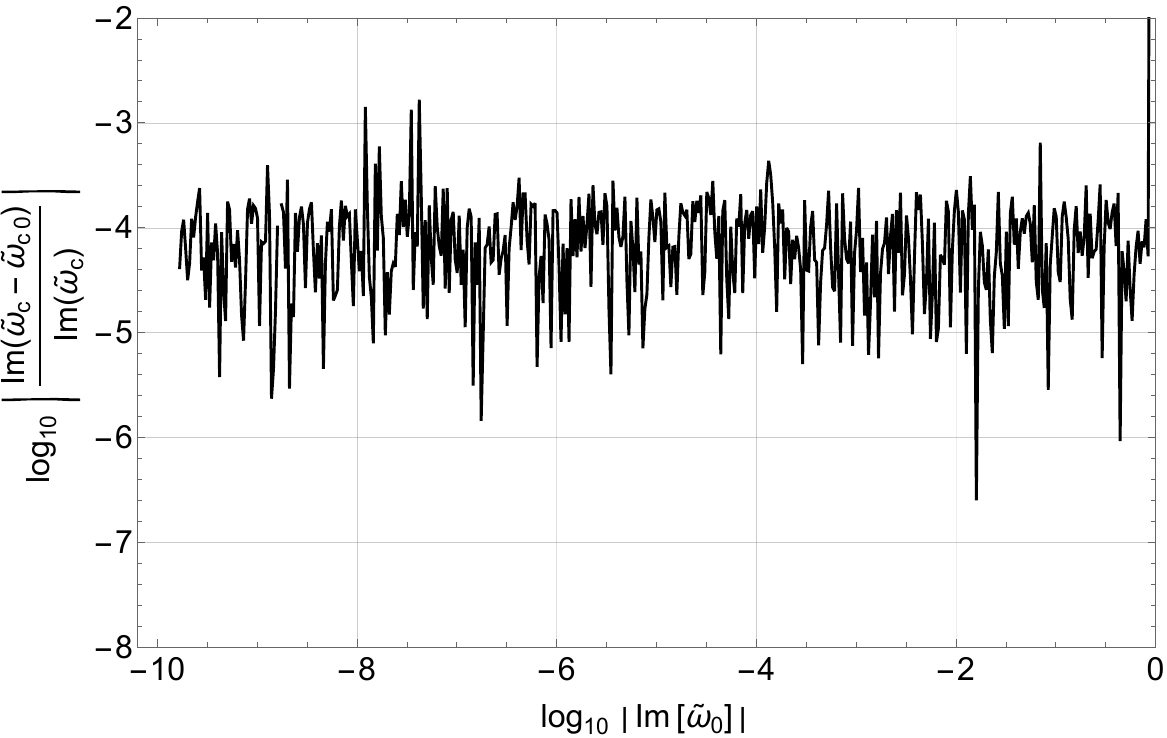}
    \caption{Coupled}
\end{subfigure}%
\begin{subfigure}[h]{0.49\linewidth}
     \centering
    \includegraphics[width=0.95\linewidth]{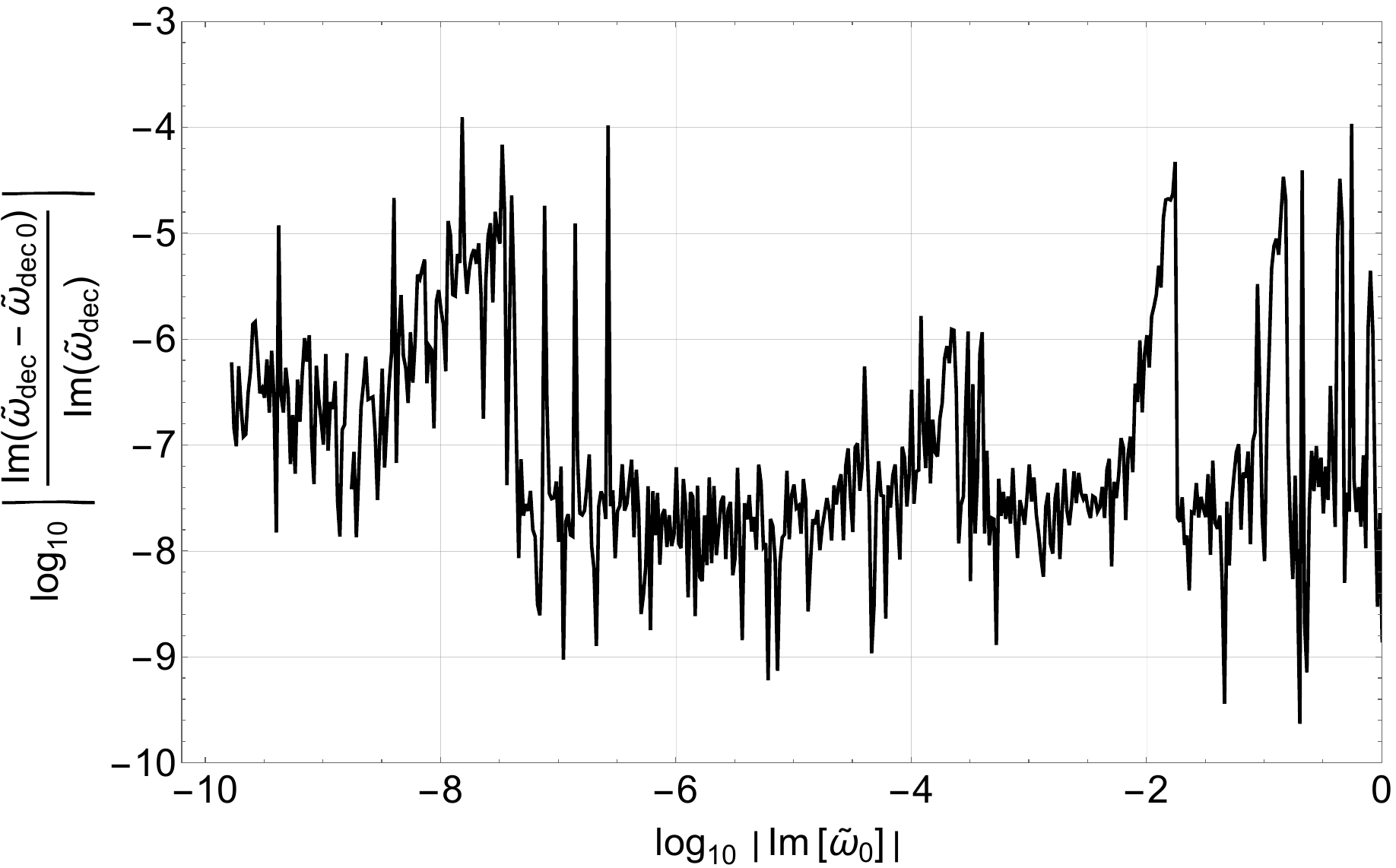}
    \caption{Decoupled}
\end{subfigure}
    \caption{Dependence on the initialization of the imaginary part $\mathrm{Im}(\tilde{\omega}_0)$ 
    of the result obtained by the root finder for the coupled system, with frequency $\tilde{\omega}_{\mathrm{c}}$, and for the decoupled system, $\tilde{\omega}_{\mathrm{dec}}$, in logarithmic scale, for the scalar-type electromagnetic polarization $\beta_+$, for $\mu=0.001$. The $\tilde{\omega}_{\mathrm{c}0}$ and
    $\tilde{\omega}_{\mathrm{dec}0}$ are the frequencies obtained by initializing the 
    root finder with the results from the isomonodromy method.}
    \label{fig:diffinit}
\end{figure}

In Fig.~\ref{fig:diffinit}, the dependence of the frequencies on the initialization of the 
root finder is analyzed, for the method involving the coupled system and the one involving the 
decoupled system. For the coupled system, the frequency $\tilde{\omega}_c$ is obtained in 
function of the imaginary part of the initializing frequency $\tilde{\omega}_0$ of the 
find root. The relative difference between $\mathrm{Im}(\tilde{\omega}_{\mathrm{c}})$ and 
$\mathrm{Im}(\tilde{\omega}_{\mathrm{c}0})$ is plotted in function of $\mathrm{Im}
(\tilde{\omega}_{0})$ in logarithmic scale, where $\mathrm{Im}(\tilde{\omega}_{\mathrm{c}0})$ 
is the imaginary part of the 
frequency obtained by initializing the root finder with the result coming from the isomonodromy 
method. The same is done for the decoupled system, where the subscript $\mathrm{dec}$ is used 
instead of $\mathrm{c}$. It is shown that the mean difference is around $10^{-4}$ for the coupled system and around 
$10^{-7}$ for the decoupled system. This clearly shows that the 
root finder with the decoupled system is more accurate than 
the root finder with the decoupled system. Furthermore, 
the root finder with the coupled system has much more difficulty to find the imaginary part of the frequency for 
$\mu < 0.001$.

\bibliography{biblio}
\bibliographystyle{JHEP}

\end{document}